\documentclass[sn-mathphys-num]{sn-jnl}% Math and Physical Sciences Numbered Reference Style 
\usepackage{graphicx}%
\usepackage{multirow}%
\usepackage{amsmath,amssymb,amsfonts}%
\usepackage{amsthm}%
\usepackage{mathrsfs}%
\usepackage[title]{appendix}%
\usepackage{textcomp}%
\usepackage{manyfoot}%
\usepackage{booktabs}%
\usepackage{listings}%
\usepackage[dvipsnames]{xcolor}
\usepackage{setspace}
\usepackage{float}
\usepackage{subcaption}
\usepackage{bm}
\usepackage{comment}
\usepackage[linesnumbered,ruled,vlined]{algorithm2e}
\newcommand{\ddd}{\mathrm{d}}
\theoremstyle{thmstyleone}%
\theoremstyle{thmstyletwo}%

\theoremstyle{thmstylethree}%

\begin{document}

\title[Article Title]{Nonlinear estimation in turbulent channel flows}

%%=============================================================%%
%% GivenName	-> \fnm{Joergen W.}
%% Particle	-> \spfx{van der} -> surname prefix
%% FamilyName	-> \sur{Ploeg}
%% Suffix	-> \sfx{IV}
%% \author*[1,2]{\fnm{Joergen W.} \spfx{van der} \sur{Ploeg} 
%%  \sfx{IV}}\email{iauthor@gmail.com}
%%=============================================================%%

\author*[1]{\fnm{Jitong} \sur{Ding}}\email{jitongd@student.unimelb.edu.au}

\author[1]{\fnm{Simon J.} \sur{Illingworth}}

\affil*[1]{\orgdiv{Department of Mechanical Engineering}, \orgname{University of Melbourne}, \orgaddress{\city{Parkville}, \postcode{3010}, \state{VIC}, \country{Australia}}}

%%==================================%%
%% Sample for unstructured abstract %%
%%==================================%%

\abstract{We design a nonlinear estimator for channel flows at $Re_{\tau}=180$ and $590$.
The nonlinear estimator uses a linear estimator structure based on the linearised Navier-Stokes equations and explicitly calculates the nonlinear forcing from the estimated velocities in physical space. 
The goal is to use the velocities at one wall-normal height to estimate the velocities at other wall-normal heights. 
The estimation performance is compared among the nonlinear estimator, the linear estimator and the linear estimator augmented with eddy viscosity. 
At $Re_{\tau}=180$, the nonlinear estimator and the linear estimator augmented with eddy viscosity outperform the linear estimator in terms of estimating the velocity magnitudes, structures and energy transfer (production and dissipation) across the channel height.
The limitations of using measurement data at one wall-normal height are discussed. 
At $Re_{\tau}=590$, the nonlinear estimator does not work well with only one measurement plane, whereas the linear estimator augmented with eddy viscosity performs well.
The performance of the nonlinear estimator at $Re_{\tau}=590$ is significantly enhanced by providing multiple measurement planes.
}


%%\pacs[JEL Classification]{D8, H51}

%%\pacs[MSC Classification]{35A01, 65L10, 65L12, 65L20, 65L70}

\maketitle

\section{Introduction}\label{sec1}
Flow estimation has a wide range of applications ranging from large-scale weather prediction \citep{richardson1922weather, le1986variational} to micro-scale flow estimation in electronic chips \citep{fan2015nanoscale}. 
However, practical limitations hinder us from accessing the full information of a fluid flow.
For example, particle image velocimetry (PIV) only measures a limited number of 2-D planes in a 3-D flow field and few materials could be used to measure the temperature inside a combustion chamber. 
The goal of flow estimation is to predict unknown quantities using limited flow measurement data. 
To achieve this, one way is to build mathematical models based on the flow physics \citep{rowley2017model, taira2017modal, taira2020modal}.
As the Navier-Stokes equations governing the flow dynamics are nonlinear and pose significant challenges, a simple way is to use the Navier-Stokes equations linearised around the mean velocity field.  

The linearised Navier-Stokes equations can explain important physical mechanisms in wall-bounded flows, such as transient energy growth and the lift-up mechanism caused by non-normality in the presence of the mean shear \citep{ellingsen1975stability,landahl1980note,trefethen1993hydrodynamic,schmid2002stability}. 
The linearised Navier-Stokes equations have been used in both laminar flows \citep{gustavsson1991energy, butler1992three, reddy1993energy, jovanovic2005componentwise} and fully developed turbulent flows \citep{del2006linear,pujals2009note,hwang2010amplification, hwang2010linear} for the fluctuation velocities following a Reynolds decomposition. 
More recently, a novel approach proposed by \citet{mckeon2010critical}
does not assume small fluctuations to the linearised Navier-Stokes equations around the mean velocity profile in fully developed turbulent flows. 
From an input-output perspective, the nonlinear forcing is treated as an unknown input, and the leading output mode of the velocity field represents the structure that is most amplified by linear mechanisms.
This linear method can extract low-rank features in fully developed turbulent flows and sheds light on reduced-order modelling \citep{mckeon2010critical,sharma2013coherent, mckeon2017engine}. 

However, resolvent analysis implicitly assumes that the nonlinear forcing is white in space and time and does not model the `shape' of the nonlinear forcing.
This can cause problems in situations where the nonlinear forcing has a significant projection onto the suboptimal input modes \citep{symon2018non, rosenberg2019role, morra2021colour}. 
It has been shown that the nonlinear forcing is structured \citep{chevalier2006state,morra2021colour,nogueira2021forcing} and embedding the knowledge of the nonlinear forcing into the linear resolvent model outperforms the pure linear resolvent model in terms of recovering flow statistics and structures \citep{zare2017colour, illingworth2018estimating,madhusudanan2019coherent,towne2020resolvent,amaral2021resolvent,fan2024eddy}.

This paper investigates channel flow estimation at $Re_{\tau}=180$ and 590. 
Similar to \citet{illingworth2018estimating}, we use a resolvent-based linear estimator with consideration of the nonlinear forcing.
The goal is to use the velocity data coming from direct numeric simulations (DNS) at one wall-normal height to estimate the velocity field at other wall-normal heights.
The key feature in this study is that we explicitly calculate the nonlinear forcing from the estimated velocities in physical space, resembling closing the loop of the resolvent-based linear estimator.
Estimation in this study is implemented using a filtering technique (Kalman filter) which involves predicting the flow states by marching the governing equations with measurement data \citep{hoepffner2005state,chevalier2006state,colburn2011state,illingworth2018estimating, oehler2018linear, gong2021modelling}. 
%Another estimation technique is smoothing which involves using time series data to search for the optimal initial condition \citep{le1986variational,wang2021state,wang2022observable}.
We compare the estimation performance between the nonlinear estimator designed in this study, the linear estimator and the linear estimator augmented with an eddy viscosity \citep{illingworth2018estimating}.
In addition, we discuss the limitations of using velocity measurement data at only one wall-normal height. 
Having looked at estimation using only one measurement plane, we proceed to explore estimation using multiple measurement planes distributed in the wall-normal direction at $Re_{\tau}=180$ and $590$.

The nonlinear estimator in this study shares some similarities with large-eddy simulation (LES). 
The main similarity is that both the nonlinear estimator and LES only resolve the large scales (the reasons for the nonlinear estimator only considering the large scales are described in \S\ref{chapter6_section_linear_estimator}).
The main difference is that the nonlinear estimator requires continuous external information from measurement data, whereas LES is an autonomous simulation that does not need external input apart from the initial condition. 

This paper is organised as follows. 
In \S\ref{chapter6_section_model_descriptions}, we revisit the linear estimator \citep{illingworth2018estimating} and introduce the nonlinear estimator. 
The DNS datasets are described in \S\ref{section_flow_descriptions}. 
Then, we proceed to the discussion of the results. 
We compare the nonlinear estimator, linear estimator and linear estimator augmented with eddy viscosity at $Re_{\tau}=180$ and $590$.
Specifically, \S\ref{chapter6_section_results_one_plane} discusses the estimation with one measurement plane, and \S\ref{chapter6_section_results_multiple_planes} discusses the estimation with multiple measurement planes. 
Conclusions are drawn in \S\ref{chapter6_section_conclusions}.

\section{Model descriptions} \label{chapter6_section_model_descriptions}
\subsection{Linear model}
Consider the non-dimensional incompressible Navier-Stokes equations for the fluctuation velocities after a Reynolds decomposition, $\mathcal{U}_i = U_i + u_i$, where $\mathcal{U}_i$, $U_i$ and $u_i$ represent the instantaneous velocity, time-averaged velocity and fluctuation velocity, respectively: 
\begin{subequations}
\begin{align}
\frac{\partial u_i}{\partial t} &=  
- u_j\frac{\partial U_i}{\partial x_j} 
- U_j\frac{\partial u_i}{\partial x_j} 
-\frac{\partial p}{\partial x_i} 
+ \frac{1}{Re_{\tau}} \frac{\partial^2 u_i}{\partial x_j \partial x_j}
+ f_i \\ 
\frac{\partial u_i}{\partial x_i} &= 0  
\end{align}
\label{eqa_chapter6_NSEs}%
\end{subequations}
where the indices $i=1,2,3$ represent the $x$ (streamwise), $y$ (spanwise) and $z$ (wall-normal) directions. 
The corresponding velocity components are denoted by $u$, $v$ and $w$. 
Pressure is denoted as $p$. 
Length scales are non-dimensionalised using the channel half-height $h$, time scales are non-dimensionalised using $h/u_{\tau}$ and pressure is non-dimensionalised using $\rho u_{\tau}^2$, where $u_{\tau}=\sqrt{\tau_w/\rho}$, $\rho$ is the density, $\tau_w$ is the mean wall shear stress and $u_{\tau}$ is the friction velocity. 
Then, the friction Reynolds number, $Re_{\tau} = hu_{\tau}/\nu$, is defined using $h$, $u_{\tau}$ and the kinematic viscosity, $\nu$. 
$U_i  = (U(z),0,0)$ represents the mean velocity profile, $f_i = - u_j\frac{\partial u_i}{\partial x_j} 
+ \overline{u_j\frac{\partial u_i}{\partial x_j}}$ represents the nonlinear forcing term including the Reynolds stresses. 
An overbar denotes the time-averaging operator.

Equations \eqref{eqa_chapter6_NSEs} can be rearranged to arrive at the Orr-Sommerfeld \& Squire equations for the wall-normal velocity $w$ and wall-normal vorticity $\eta = \frac{\partial u}{\partial y} - \frac{\partial v}{\partial x}$. 
Due to the homogeneity in the streamwise and spanwise directions for channel flows, the two-dimensional Fourier transform in these two directions can be applied with the Chebyshev distribution in the wall-normal direction to arrive at the state-space model at streamwise wavenumber $k_x$ and spanwise wavenumber $k_y$ \citep{schmid2002stability}:
\begin{subequations}
\begin{align}
\frac{\ddd }{\ddd t}
\hat{\bm{x}}(t)&=
\mathcal{A} \hat{\bm{x}}(t) +
\mathcal{B} \hat{\bm{f}}(t)
\\ 
\hat{\bm{y}}&=
\mathcal{C} \hat{\bm{x}}(t)
\end{align}
\label{eqa_state_space}%
\end{subequations}
where $\hat{\bm{x}}=[\hat{\bm{w}}\;\;\hat{\bm{\eta}}]^T$; $\hat{\bm{y}}=[\hat{\bm{u}}\;\;\hat{\bm{v}}\;\;\hat{\bm{w}}]^T$ and $\hat{\bm{f}}=[\hat{\bm{f}}_x\;\;\hat{\bm{f}}_y\;\;\hat{\bm{f}}_z]^T$. 
Variables with $\hat{\;}\;$ indicate that those variables are in Fourier space.
The detailed information of \eqref{eqa_state_space} is described in Appendix \ref{chapter6_section_appendix_OSS_state_space_model}. 

\subsection{Linear estimator} \label{chapter6_section_linear_estimator}
The goal is to use limited velocity measurement data to estimate the velocities at other locations. 
To achieve this, we use a Kalman filter based on the linear model \eqref{eqa_state_space}.
Consider a linear state-space model with zero input:
\begin{subequations}
\begin{align} 
\frac{\ddd }{\ddd t}
\hat{\bm{x}}(t)&=
\mathcal{A} \hat{\bm{x}}(t) +
\mathcal{B} \hat{\bm{d}}(t) \label{eqa_system_dynamics}
\\
\hat{\bm{y}}_{mea}(t) &=
\mathcal{C}_{mea} \hat{\bm{x}}(t) + \hat{\bm{n}}(t) \label{eqa_measurement}
\end{align}
\label{eqa_state_space_KF}%
\end{subequations}
where $\hat{\bm{d}}$ is the system process noise vector and $\hat{\bm{n}}$ is the measurement noise vector. 
There are two main differences between \eqref{eqa_state_space} and \eqref{eqa_state_space_KF}.
First, $\hat{\bm{d}}(t)$ in \eqref{eqa_state_space_KF} is Gaussian white noise whereas $\hat{\bm{f}}(t)$ in \eqref{eqa_state_space} is the nonlinear forcing. 
To comply with the Kalman filter setting, we treat the nonlinear forcing as white noise. 
Second, the output $\hat{\bm{y}}_{mea}(t)$ in \eqref{eqa_state_space_KF} only consists of a small portion of the full state $\hat{\bm{x}}(t)$, corresponding to the estimation problem that we have the velocity measurement at only one wall-normal height among the full velocity states across the channel height. 
The extraction of the measurement data at a specific wall-normal height is implemented using Barycentric Lagrange interpolation \citep{berrut2004barycentric}.
If the state-space model \eqref{eqa_state_space_KF} has $n$ states, $p$ disturbances and $q$ measurements, then the corresponding matrices have the following dimensions: $\mathcal{A} \in \mathbb{C}^{n \times n}$; $\mathcal{B} \in \mathbb{C}^{n \times p}$; and $\mathcal{C}_{mea} \in \mathbb{C}^{q \times n}$.

The goal is to estimate the full state $\hat{\bm{x}}(t)$ according to the dynamics of the system \eqref{eqa_system_dynamics} and the measurement data \eqref{eqa_measurement}. 
The equation of the estimated full state $\hat{\bm{x}}_{est}(t)$ is \citep{seron2012fundamental}:
\begin{equation}
\frac{\ddd}{\ddd t} \hat{\bm{x}}_{est} = \mathcal{A} \hat{\bm{x}}_{est}(t) +
\mathcal{L} [\hat{\bm{y}}_{mea}(t) - \mathcal{C}_{mea} \hat{\bm{x}}_{est}(t)]
\label{eqa_estimated_state}
\end{equation}
where $\mathcal{L}$ is the Kalman filter gain obtained by solving a Riccati equation to minimise the difference between the estimated state and true state $\lVert \hat{\bm{x}}_{est} (t) - \hat{\bm{x}}(t) \rVert^2$.
Measurement data either from experiments or DNS is supplied at a measurement sampling time $\Delta T_{m}$. 
Therefore, the model in the continuous time domain \eqref{eqa_state_space_KF} is transformed into the discrete time domain.
The block diagram for the linear estimator is shown in figure \ref{fig_simple_linear_estimator}.
Further details are given in Appendix \ref{chapter6_section_appendix_linear_estimator}.

\begin{figure}[H]
\centering
\includegraphics[scale=0.30]{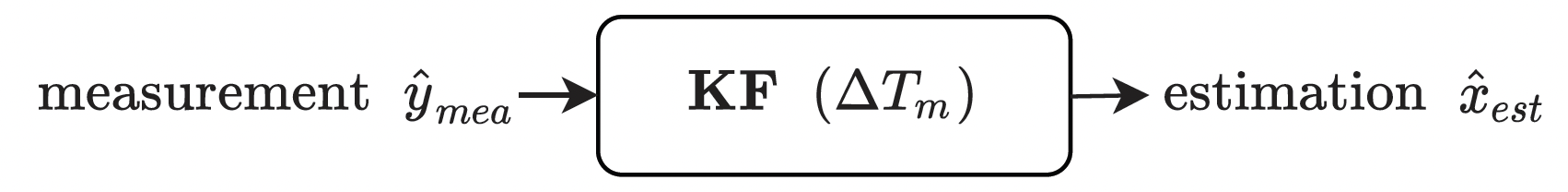}
\caption[The diagram of the linear estimator. ]
{
Block diagram of the linear estimator. 
`KF' represents Kalman filter which is implemented at measurement sampling time $\Delta T_m$. 
The input is the measurement $\hat{y}_{mea}$ and the output is the estimate $\hat{x}_{est}$, according to \eqref{eqa_estimated_state}. 
}
\label{fig_simple_linear_estimator}
\end{figure}

The Kalman filter described above is implemented at a single wavenumber pair $(k_x,k_y)$ in Fourier space, so that we form a separate Kalman filter for each wavenumber pair of interest. 
The same estimator could be applied at other wavenumber pairs.
We choose to consider the energetically dominant wavenumber pairs corresponding to the large scales \citep{illingworth2018estimating, amaral2021resolvent,arun2023towards}. 
There are three reasons.
First, large scales contain the most kinetic energy and further increasing the wavenumber pairs does not change the estimation results significantly.
Second, considering a small number of wavenumber pairs reduces computational cost and storage, achieving the goal of reduced-order modelling.
Third, we do not expect to obtain accurate estimation of small structures because they are not coherent across the channel height \citep{madhusudanan2019coherent}.

\subsection{Nonlinear estimator}
The idea of designing the nonlinear estimator is to explicitly calculate the nonlinear forcing from the estimated velocities, closing the loop of the resolvent-based linear estimator.
We consider a linear state-space model with input $\hat{\bm{f}}(t)$.
\begin{subequations}
\begin{align} 
\frac{\ddd }{\ddd t}
\hat{\bm{x}}(t)&=
\mathcal{A} \hat{\bm{x}}(t) +
\mathcal{B} \hat{\bm{d}}(t) +  
\mathcal{B} \hat{\bm{f}}(t)
\label{eqa_system_dynamics_for_nonlinear}
\\
\hat{\bm{y}}_{mea}(t) &=
\mathcal{C}_{mea} \hat{\bm{x}}(t) + \hat{\bm{n}}(t) \label{eqa_measurement_for_nonlinear}
\end{align}
\label{eqa_state_space_KF_for_nonlinear}%
\end{subequations}
Different from the linear state-space model with zero input \eqref{eqa_state_space_KF} in which we assume the disturbance $\hat{\bm{d}}(t)$ as unstructured nonlinear forcing, the input term $\hat{\bm{f}}(t)$ is the nonlinear forcing in the linear model \eqref{eqa_state_space_KF_for_nonlinear}.
The equation of the estimated state $\hat{\bm{x}}_{est}$ is then \citep{seron2012fundamental}:
\begin{equation}
\frac{\ddd}{\ddd t} \hat{\bm{x}}_{est} = \mathcal{A} \hat{\bm{x}}_{est}(t) +
\mathcal{B} \hat{\bm{f}}(t) + 
\mathcal{L} [\hat{\bm{y}}_{mea}(t) - \mathcal{C}_{mea} \hat{\bm{x}}_{est}(t)]
\label{eqa_estimated_state_for_nonlinear}
\end{equation}

We calculate the nonlinear forcing using the nonlinear dynamics $f_i = -u_j \frac{\partial u_i}{\partial x_j}$ in physical space, where the velocities $u_i$ are obtained from the estimated velocities $\hat{x}_{est}$ in \eqref{eqa_estimated_state_for_nonlinear} using the inverse Fourier transform.
With this idea, we come to the structure of the nonlinear estimator as illustrated in figure \ref{fig_simple_nonlinear_estimator1}.
Further details are given in Appendix \ref{chapter6_section_appendix_nonlinear_estimator}.

\begin{figure}[H]
\centering
\includegraphics[scale=0.30]{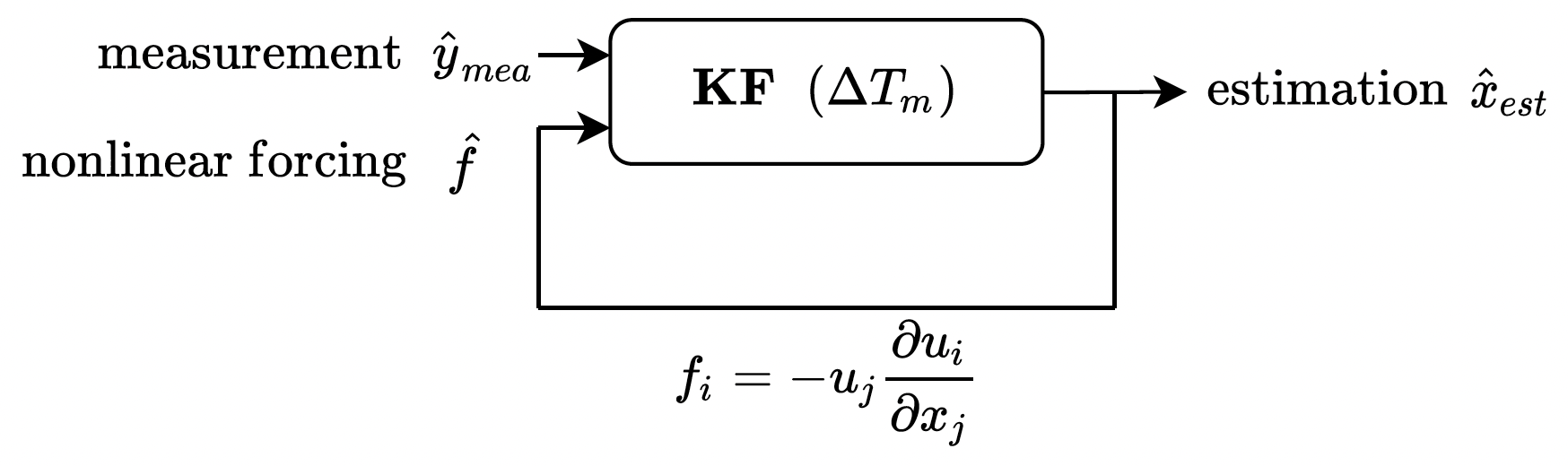}
\caption[The diagram of the nonlinear estimator. ]
{
Block diagram of the nonlinear estimator. 
`KF' represents Kalman filter which is implemented at measurement sampling time $\Delta T_m$. 
The two inputs are the measurement $\hat{y}_{mea}$ and the nonlinear forcing $\hat{f}$;
the output is the estimation $\hat{x}_{est}$, according to \eqref{eqa_estimated_state_for_nonlinear}. 
The nonlinear forcing is obtained from the estimated velocities using $f_i = -u_j \frac{\partial u_i}{\partial x_j}$. 
}
\label{fig_simple_nonlinear_estimator1}
\end{figure}

\begin{figure}[H]
\centering
\includegraphics[scale=0.36]{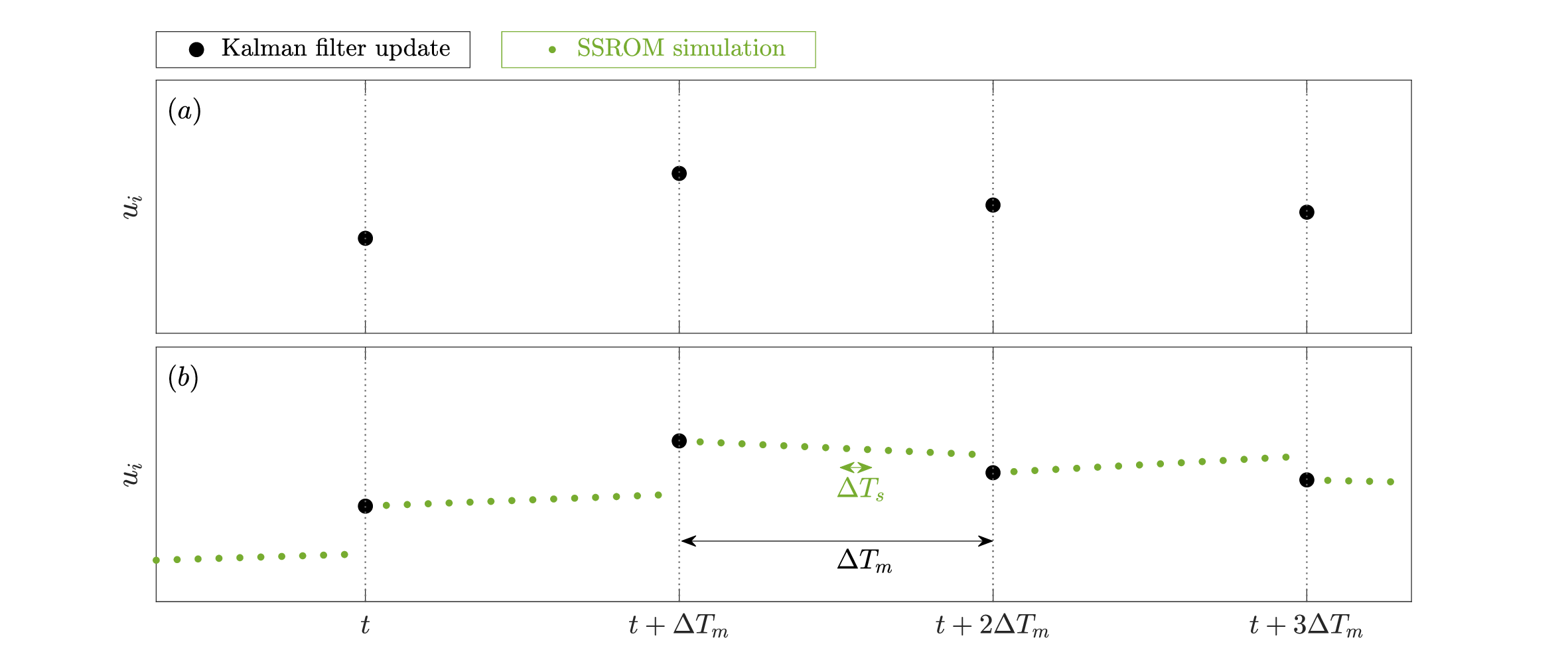}
\caption[An illustration of the estimation process.]
{An illustration of the estimation process:
(a) the nonlinear estimator in figure \ref{fig_simple_nonlinear_estimator1};
(b) the revised nonlinear estimator in figure \ref{fig_simple_nonlinear_estimator2}.
}
\label{fig_Ts_Tm}
\end{figure}

We tried the nonlinear estimator in figure \ref{fig_simple_nonlinear_estimator1} with measurement data provided every measurement sampling time step $\Delta T_m$, as illustrated in figure \ref{fig_Ts_Tm}(a). 
The results were all divergent even when measurement data at every wall-normal height were provided. 
Thus, the divergent problem is not primarily related to insufficient measurement data. 
So, we switched the attention to another variable crucial to the nonlinear estimation: the measurement sampling time $\Delta T_m$ which is also the marching time step for the Kalman filter.

One straightforward solution is to use high time-resolution DNS data with a much smaller measurement sampling time. 
However, at that time we did not have access to high time-resolution DNS data so the measurement sampling time $\Delta T_m$ had to be fixed. 
This also could be the case in an experiment (for example PIV) or even in a DNS that we could not obtain velocity measurements with high time-resolution due to technical constraints.
Nevertheless, we could still decrease the marching time step for the Kalman filter state prediction from $\Delta T_m$ to $\Delta T_s$, where $\Delta T_s \ll \Delta T_m$.
In this case, at time $t$, we have the estimated velocities $\hat{x}_{est}|_{t}$;
at time $t+\Delta T_m$ (because we do not change the measurement sampling time $\Delta T_m$), the Kalman filter needs to give estimation according to the measurement at time $t+\Delta T_m$ and the previous nonlinear forcing at time 
$t+\Delta T_m - \Delta T_s$ with the marching time step of $\Delta T_s$.
One question arises: how do we go from the estimated velocities $\hat{x}_{est}|_{t}$ at time $t$ to the nonlinear forcing $\hat{f}|_{t+\Delta T_m - \Delta T_s}$ at time $t+\Delta T_m - \Delta T_s$?

In light of the method of introducing the nonlinear dynamics back to the estimator as shown in figure \ref{fig_simple_nonlinear_estimator1}, we combine the Orr-Sommerfeld \& Squire state-space model \eqref{eqa_state_space} with this nonlinear dynamics feedback loop.
We choose a time step $\Delta T_s$ which is much smaller than $\Delta T_m$ to discretise the state-space model \ref{eqa_state_space}.
First, we use the nonlinear forcing at time $t$ as input to drive the state-space model \eqref{eqa_state_space} to obtain the velocities at time $t+\Delta T_s$. 
Second, we use the nonlinear dynamics in physical space $f_i = -u_j \frac{\partial u_i}{\partial x_j}$ to calculate the nonlinear forcing at time $t+\Delta T_s$. 
Third, the nonlinear forcing at time $t+\Delta T_s$ serves as input to drive the state-space model \eqref{eqa_state_space} again to iterate the process until time $t+\Delta T_m - \Delta T_s$.
These three steps resemble an autonomous simulation.
We embed this autonomous simulation as shown by the green dots in figure \ref{fig_Ts_Tm}(b) between two consecutive Kalman filter updates determined by the measurement sampling time $\Delta T_m$.
To align with the estimation process, 
we consider the same wavenumber pairs used in the Kalman filter estimation for this autonomous simulation.
Since we only consider large scales in this simulation, we name this simulation as state-space reduced-order model (SSROM) simulation. 
Further details about the SSROM simulation are given in Appendix \ref{chapter6_section_appendix_SSROM}. 

The block diagram for the revised nonlinear estimator embedded with SSROM simulation is shown in figure \ref{fig_simple_nonlinear_estimator2}.
Further details are given in Appendix \ref{chapter6_section_appendix_nonlinear_estimator_SSROM}.
From the nonlinear estimator illustrated in figure \ref{fig_simple_nonlinear_estimator1} to the revised nonlinear estimator illustrated in figure \ref{fig_simple_nonlinear_estimator2}, the stability of the estimator has greatly improved. 
Comparing the nonlinear estimators in figures \ref{fig_simple_nonlinear_estimator1} and \ref{fig_simple_nonlinear_estimator2}, the main similarity is that they both explicitly calculate the nonlinear forcing rather than treat it as an unknown forcing and use the structured nonlinear forcing to serve as input to the Kalman filters.
The main difference is the marching time step for the state prediction.
The nonlinear estimator in figure \ref{fig_simple_nonlinear_estimator1} marches forward in time at the measurement sampling time $\Delta T_m$ as shown in figure \ref{fig_Ts_Tm}(a), which easily triggers divergence if $\Delta T_m$ is too big. 
The revised nonlinear estimator in figure \ref{fig_simple_nonlinear_estimator2} marches forward in time at a much smaller time step $\Delta T_s$ achieved by the inclusion of an autonomous simulation from the SSROM between two consecutive Kalman filter updates, as shown in figure \ref{fig_Ts_Tm}(b).

\begin{figure}[H]
\centering
\includegraphics[scale=0.25]{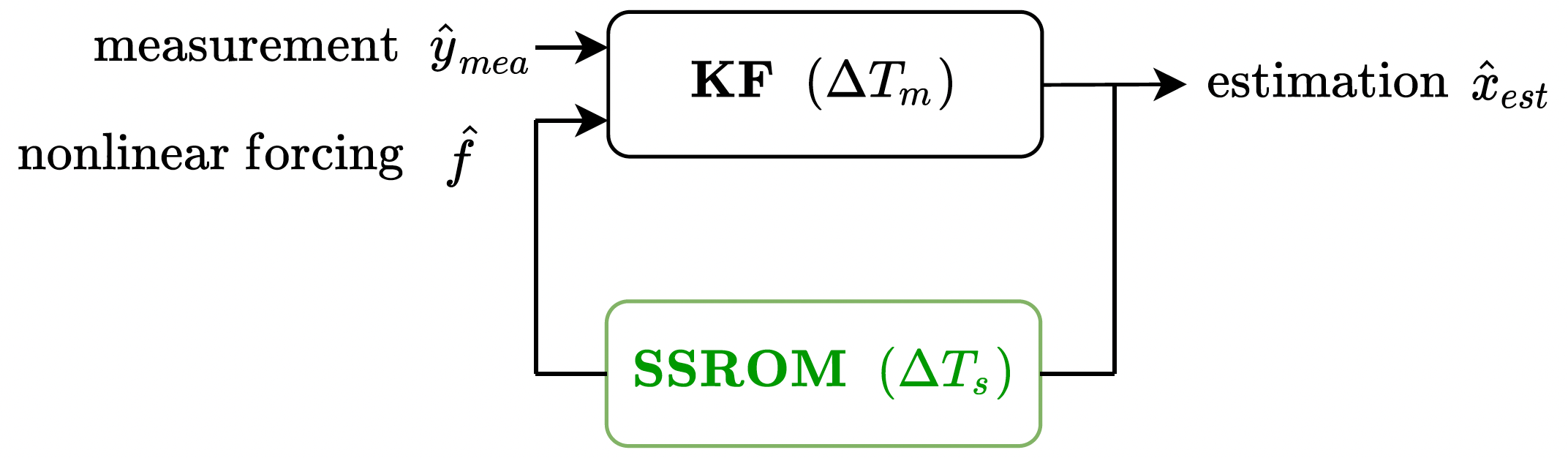}
\caption[The diagram of the revised nonlinear estimator. ]
{
Block diagram of the revised nonlinear estimator. 
`KF' represents Kalman filter which is used at every measurement sampling time $\Delta T_m$ with a marching time step of $\Delta T_s$. 
The two inputs are measurement $\hat{y}_{mea}$ and nonlinear forcing $\hat{f}$;
the output is estimation $\hat{x}_{est}$, according to \eqref{eqa_estimated_state_for_nonlinear}. 
Between two consecutive Kalman filter updates, SSROM simulation running at $\Delta T_s$ $(\Delta T_s \ll \Delta T_m)$ is used to obtain the nonlinear forcing $\hat{f}$ for the next Kalman filter update. 
}
\label{fig_simple_nonlinear_estimator2}
\end{figure}

\section{Flow descriptions} \label{section_flow_descriptions}
Direct numerical simulations are performed using a staggered-grid fourth-order finite-difference solver \citep{chung2014idealised}. 
Table \ref{table_DNS} summarises the simulation parameters. 
For $L_x = 2\pi$ and $L_y=\pi$, the maximum wavenumbers resolved by the simulation are $k_x = \pm 55,\;k_y = \pm 110$ for $Re_{\tau} = 180$ and $k_x = \pm 191,\;k_y = \pm 382$ for $Re_{\tau} = 590$;
the minimum wavenumbers are $k_x = \pm 1,\;k_y = \pm 2$ for both Reynolds numbers.

\setlength{\tabcolsep}{8pt} % Default value: 6pt
\renewcommand{\arraystretch}{1.2} % Default value: 1

\begin{table}[h]
\caption{Parameter setup of the DNS. $L$, domain length; $n$, number of grid points; $\Delta^+$, grid-spacing in viscous units; $\Delta t$ is the simulation time step.}
\label{table_DNS}
\begin{tabular}{lllllllll}
\hline
$Re_{\tau}$ & $L_x$  & $L_y$ & $n_x \times n_y \times n_z$ & $\Delta x^+$ & $\Delta y^+$ & $\Delta z^+_{max}$ & $\Delta z^+_{min}$ & $\Delta t^+$ \\ 
180         & $2\pi$ & $\pi$ & $112 \times 112 \times 150$ & 10.00        & 5.05         & 3.77               & 0.04                  & 0.36      \\ 
590         & $2\pi$ & $\pi$ & $384 \times 384 \times 500$ & 9.65        & 4.83         & 3.71               & 0.01                  & 0.12       \\
\hline
\end{tabular}
\end{table}

\begin{figure}[H]
\centering
\begin{flushleft}
\begin{subfigure}{.48\textwidth}
  \includegraphics[width=1.0\linewidth]{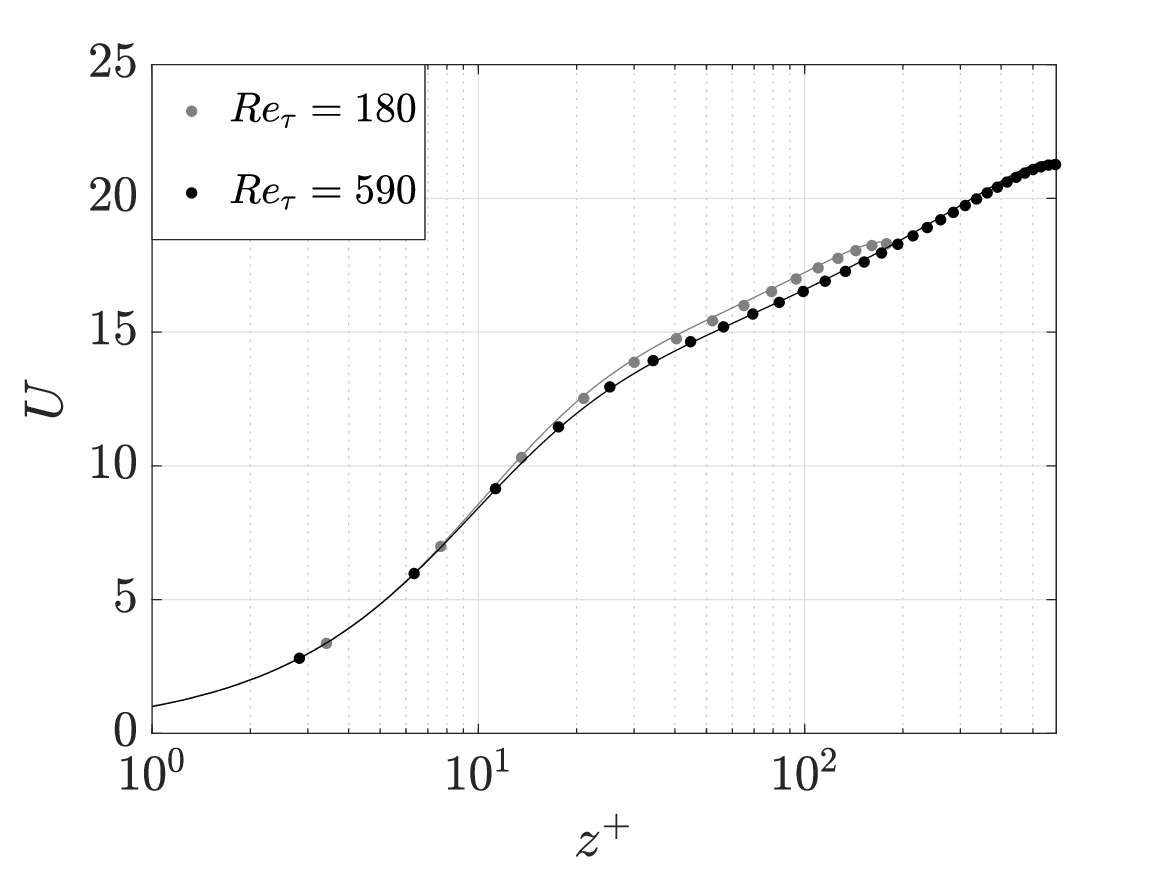}
\end{subfigure}
\begin{subfigure}{.48\textwidth}
  \centering
  \includegraphics[width=1.0\linewidth]{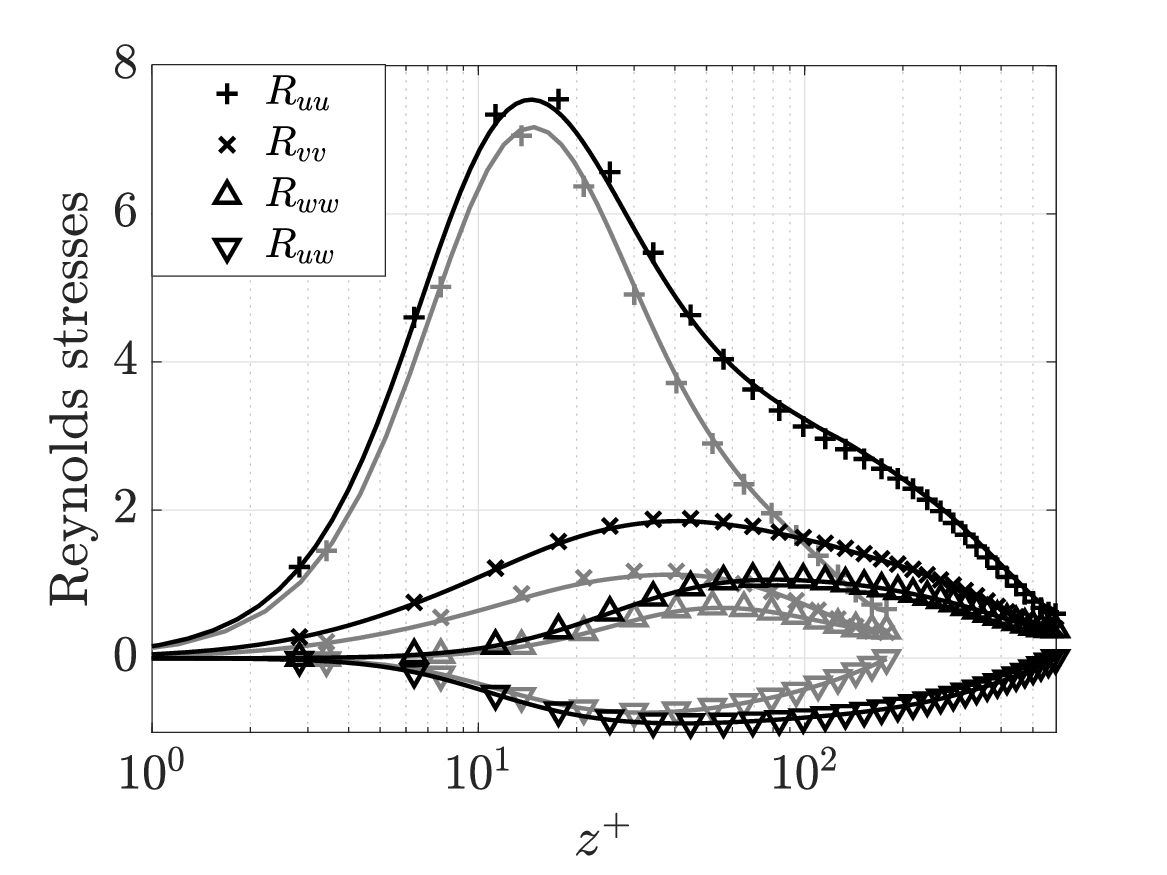}
\end{subfigure}
\end{flushleft}
\caption{Comparison between the DNS dataset (solid lines) and the standard DNS dataset \protect \citep{moser1999direct} (markers). (a) Mean streamwise velocity. (b) Reynolds stresses. 
Grey, $Re_{\tau}=180$; black, $Re_{\tau}=590$. }
\label{fig_DNS_dataset}
\end{figure}

The time-averaged first-order and second-order statistics of the present DNS data show good agreement with \citet{moser1999direct}, as shown in figure \ref{fig_DNS_dataset}.

\section{Estimation using one measurement plane} \label{chapter6_section_results_one_plane}
Now we discuss the estimation results for the different estimators introduced previously.
\S\ref{chapter6_section_NE_LE_180} and \S\ref{chapter6_section_NE_LEe_180} discuss the estimation at $Re_{\tau}=180$.
\S\ref{chapter6_section_change_measurement_location} examines the effect of changing the location of the measurement plane at $Re_{\tau}=180$. 
\S\ref{chapter6_section_estimation_590_one_plane} discusses the estimation at $Re_{\tau}=590$.
To simplify the discussion, `NE' refers to the nonlinear estimator in figure \ref{fig_simple_nonlinear_estimator2}, `LE' refers to the linear estimator in figure \ref{fig_simple_linear_estimator} and `LEe' refers to the linear estimator augmented with eddy viscosity to be introduced in \S\ref{chapter6_section_NE_LEe_180}. 

\subsection{Comparison between NE and LE at $Re_{\tau}=180$} \label{chapter6_section_NE_LE_180}
We apply the nonlinear estimator (NE) and the linear estimator (LE) at $Re_{\tau}=180$ and discuss the result comparison between these two estimators. 
The goal is to use the velocity data at a single wall-normal height to estimate the velocities at other wall-normal heights. 
Since the estimation is implemented in a channel, the measurement velocity at one wall-normal height is mirrored into the other half of the channel. 
To ensure a fair comparison, the same estimation setting is applied to NE and LE.
The wavenumbers considered at $Re_{\tau}=180$ are $\lvert k_x \rvert \leq 10$ and $\lvert k_y \rvert \leq 20$, corresponding to structures with $\lambda_x^+ \geq 113$ and $\lambda_y^+ \geq 57$.
In the wall-normal direction we use 129 Chebyshev points and convergence has been checked by doubling the number of Chebyshev points.
The measurement data (velocities) come from $z^+=15$, where the turbulent kinetic energy reaches its maximum in the wall-normal direction. 
The measurement data comes from DNS and the temporal resolution is $\Delta T_m = 0.02$ which is small enough to resolve the frequency of the smallest scale according to Taylor's hypothesis. 
The SSROM simulation time step is set to be $\Delta T_s = 0.0005$.
The estimation starts with zero initial condition. 
We present the estimation results starting from $t=2(t^+=360)$ to $t=3(t^+=540)$ to minimise the initial transient. 

\begin{figure}[H]
\centering
\includegraphics[scale=0.35]{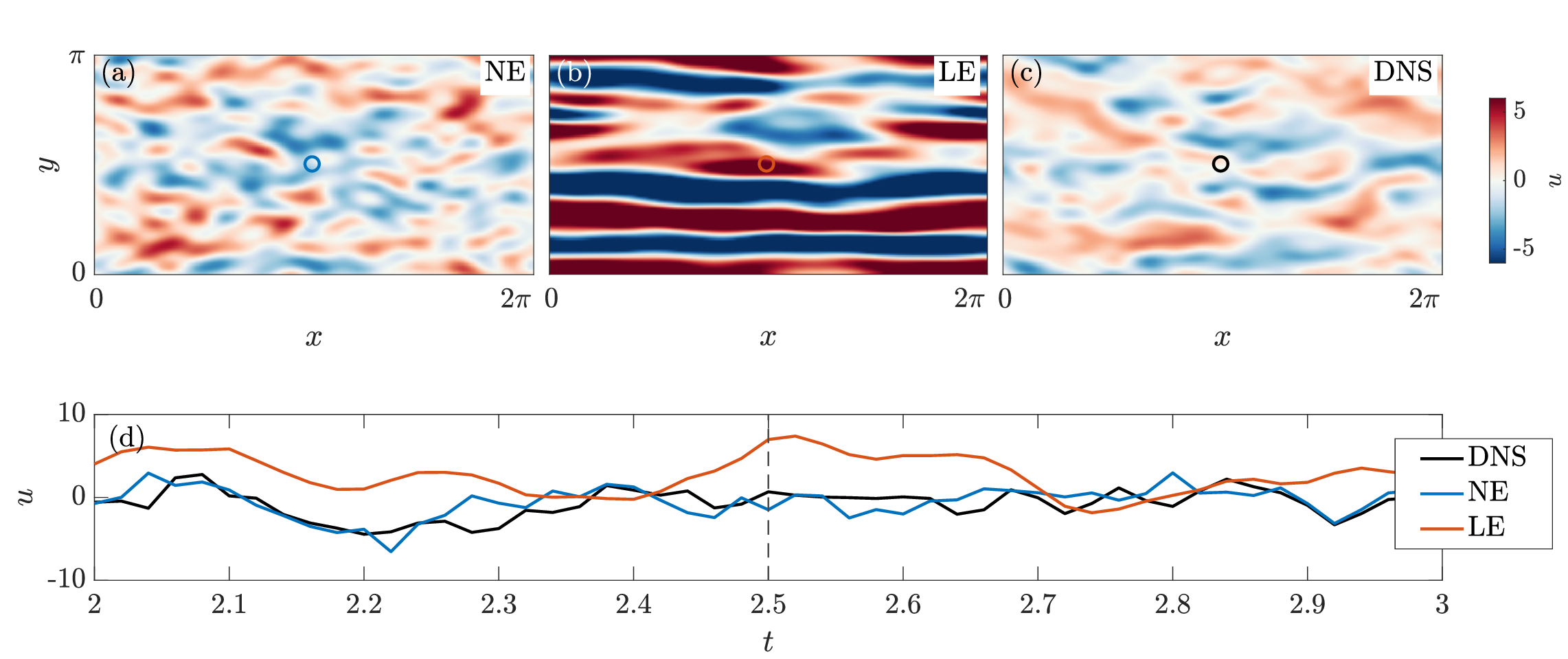}
\caption[Estimated streamwise fluctuation velocities at $z^+=50$ for NE and LE at $Re_{\tau}=180$.]
{Streamwise velocity $u$ for (a) NE, (b) LE and (c) DNS data in the $xy$ plane at $z^+=50$. 
The measurement data comes from $z^+=15$.
The circles in the middle of (a,b,c) denote the locations of the streamwise velocity plotted in (d). 
(d) Streamwise velocity varying with time at the centre of the $xy$ plane at $z^+=50$.
The vertical dashed line denotes the time instant for (a,b,c). 
}
\label{fig_z50xy_nonlin_lin}
\end{figure}

Figure \ref{fig_z50xy_nonlin_lin} shows the estimated velocity field at $z^+=50$. 
The DNS velocity field in figure \ref{fig_z50xy_nonlin_lin}(c) only contains the large scales considered in the estimation. 
The estimated velocities from NE (figure \ref{fig_z50xy_nonlin_lin}(a)) are around the similar magnitudes of the DNS data (figure \ref{fig_z50xy_nonlin_lin}(c)). 
The estimated velocities from LE (figure \ref{fig_z50xy_nonlin_lin}(b)) are much overpredicted compared with the DNS data (figure \ref{fig_z50xy_nonlin_lin}(c)), which were also observed in previous studies \citep{illingworth2018estimating, amaral2021resolvent}.
The features related to the magnitudes can also be seen in figure \ref{fig_z50xy_nonlin_lin}(d).
As for the structures, NE gives some correct large-scale structures at the corresponding locations in the DNS but it also generates small-scale structures that are non-existent in the DNS;
LE only gives the large-scale structures. 

\begin{figure}[H]
\centering
\includegraphics[scale=0.36]{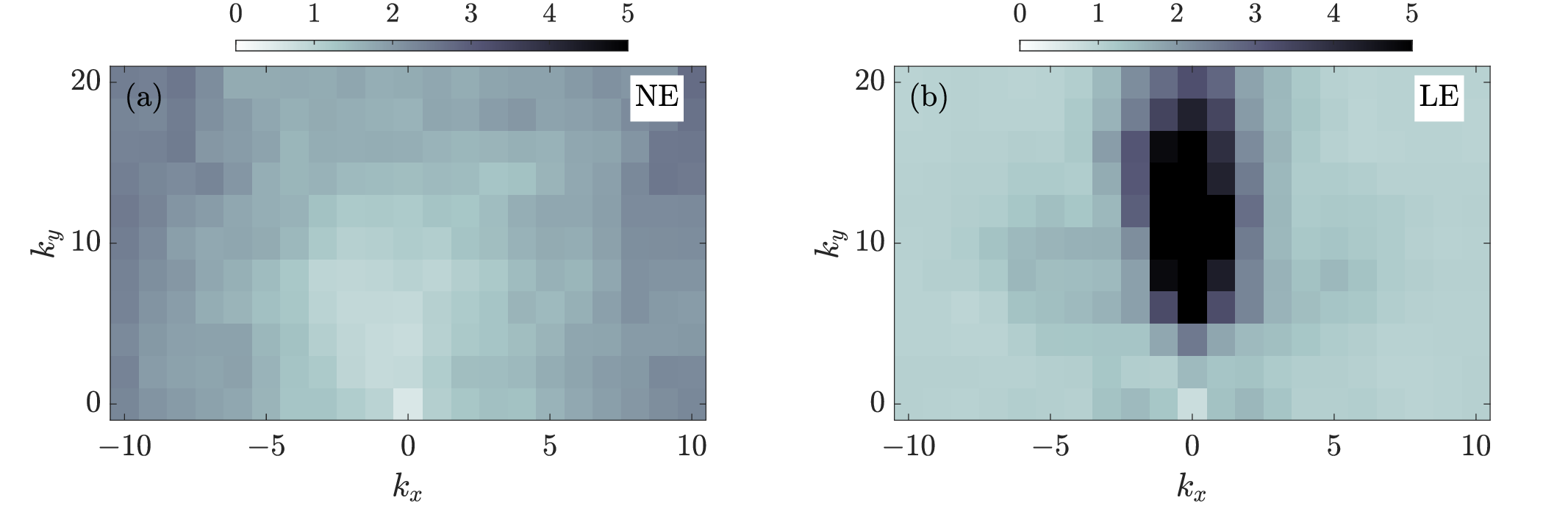}
\caption[Estimation error in Fourier space at $z^+=50$ for NE and LE at $Re_{\tau}=180$. ]
{Error in Fourier space \eqref{eqa_error_F} for 
(a) NE and 
(b) LE at $z^+=50$.
}
\label{fig_z50error_nonlin_lin}
\end{figure}

Since the estimation is conducted in Fourier space, we evaluate the estimation performance at each wavenumber pair by defining the error \citep{illingworth2018estimating}:
\begin{equation}
\hat{\epsilon}(k_x,k_y) = 
\frac
{\sqrt{\int \lvert \hat{u}_{est} - \hat{u}_{DNS} \rvert^2   \ddd t}}
{\sqrt{\int \lvert \hat{u}_{DNS} \rvert^2   \ddd t}}
\label{eqa_error_F}
\end{equation}

Figure \ref{fig_z50error_nonlin_lin} shows the estimation error in Fourier space for NE and LE. 
In figure \ref{fig_z50error_nonlin_lin}(a), we see that NE performs better for large streamwise scales with small $k_x$ than small streamwise scales with large $k_x$. 
This corresponds to the observation of incorrect small scales from NE (\ref{fig_z50xy_nonlin_lin}(a)).
In figure \ref{fig_z50error_nonlin_lin}(b), we see that LE has substantial errors for the streamwise-elongated scales ($k_x=0$).
This corresponds to the observation of large magnitudes of the large streamwise scales from LE (\ref{fig_z50xy_nonlin_lin}(b)), which is also noted in the previous study that the error from the linear estimator is large for streamwise-elongated scales \citep{illingworth2018estimating}.
Comparing figures \ref{fig_z50error_nonlin_lin}(a,b), we see that NE performs better than LE at $z^+=50$.

\begin{figure}[H]
\centering
\includegraphics[scale=0.36]{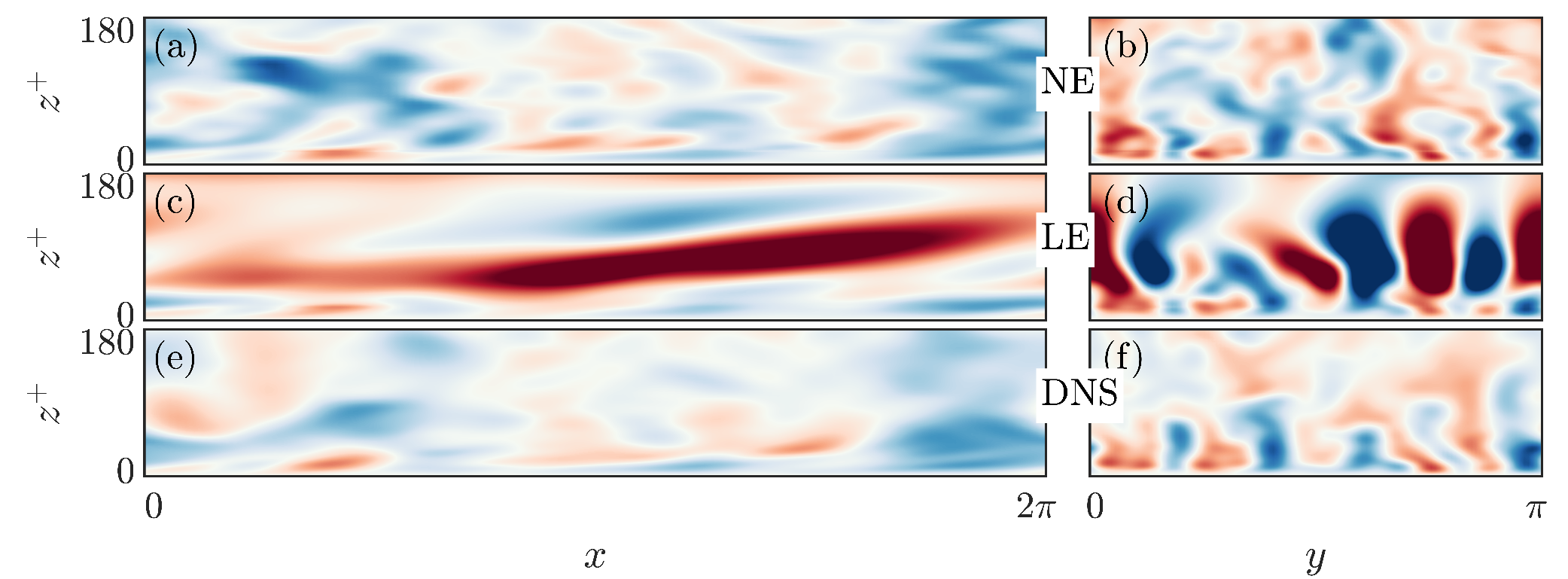}
\caption[2-D estimated streamwise velocities for NE and LE in the $xz$ plane and $yz$ plane at $Re_{\tau}=180$.]
{(a,c,e) Streamwise velocity field in the $xz$ plane at $y=\frac{\pi}{2}$.
(b,d,f) Streamwise velocity field in the $yz$ plane at $x=\pi$. 
(a,b) NE;
(c,d) LE and 
(e,f) the DNS data.
The limits of the colour map are the same as that in figure \ref{fig_xzyz_nonlinear_lin}.
}
\label{fig_xzyz_nonlinear_lin}
\end{figure}

Having looked at the estimated velocity field at $z^+=50$, we now look at the velocity field across the wall-normal height at a particular instant in time, as shown in figure \ref{fig_xzyz_nonlinear_lin}. 
We see that NE gives the correct shapes of the near-wall structures since the measurement plane is at $z^+=15$ and the shapes of the estimated velocity field far from the wall do not match the DNS (figures \ref{fig_xzyz_nonlinear_lin}(a,b,e,f)). 
LE overpredicts the velocity magnitudes as observed from the colour scale and the shapes of the estimated velocity field do not match the DNS (figures \ref{fig_xzyz_nonlinear_lin}(c,d,e,f)).

We quantify the estimation performance across the wall-normal height by defining the error and correlation \citep{chevalier2006state,colburn2011state}:
\begin{subequations}
\begin{align}
\epsilon(z) &=  \frac{\sqrt{\iiint \; (u_{est}-u_{DNS})^2 \;\ddd x \;\ddd y \;\ddd t}}{\sqrt{\iiint \; u_{DNS}^2 \;\ddd x \;\ddd y \;\ddd t}} \label{eqa_error_z} \\ 
\text{corr}(z) &= \frac{\iiint \; u_{est} u_{DNS} \;\ddd x \;\ddd y \;\ddd t}{\sqrt{\iiint \; u_{est}^2 \;\ddd x \;\ddd y \;\ddd t} \sqrt{\iiint u_{DNS}^2 \;\ddd x \;\ddd y \;\ddd t}}  \label{eqa_corr_z}%
\end{align}
\end{subequations}

\begin{figure}[H]
\centering
\includegraphics[scale=0.37]{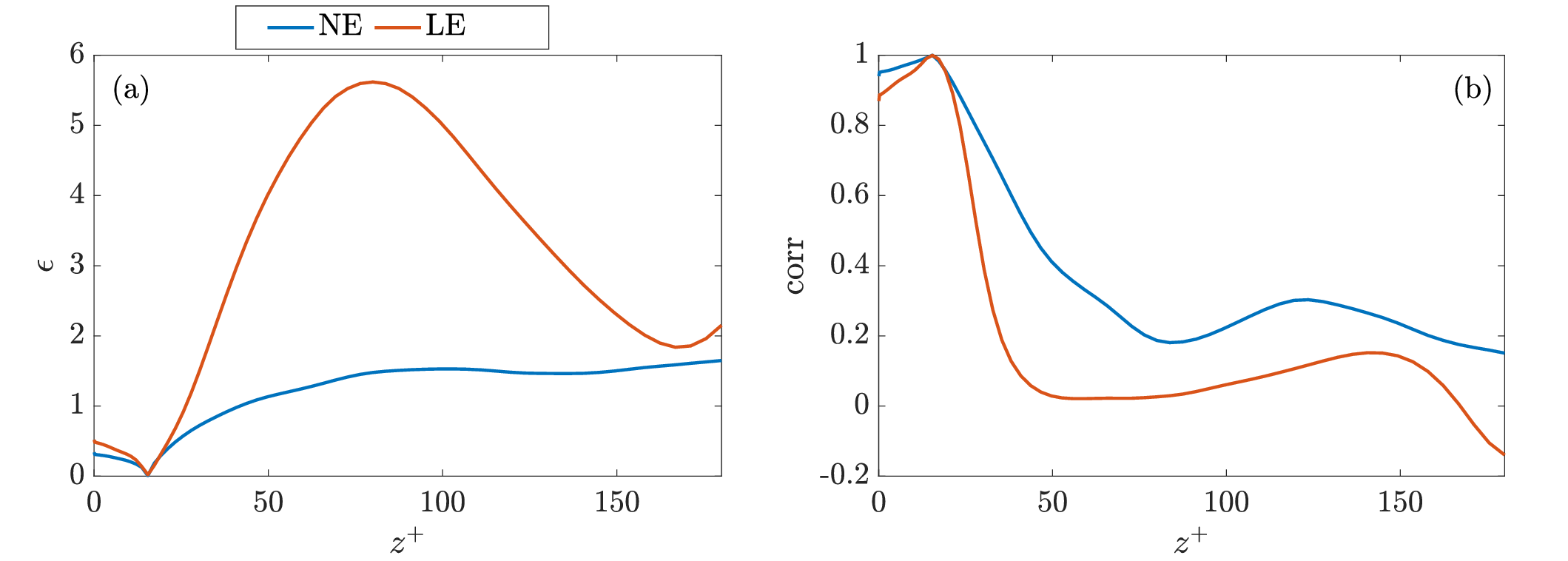}
\caption[Estimation error and correlation across the wall-normal height for NE and LE.]
{The error \eqref{eqa_error_z} and correlation \eqref{eqa_corr_z} across the wall-normal height for NE and LE. 
}
\label{fig_error_corr_nonlin_lin}
\end{figure}

Equation \eqref{eqa_error_z} quantifies the deviation between the estimated quantity and true quantity (DNS data).
Correlation \eqref{eqa_corr_z} quantifies the phase alignment between the estimated quantity and true quantity.
Figure \ref{fig_error_corr_nonlin_lin} shows the errors and correlations for NE and LE.
An indicator of good estimation performance is a low error approaching zero and a high correlation approaching one. 
We see that both estimators perform the best at $z^+=15$ where the measurement data is provided, and the estimation performance deteriorates away from the measurement plane. 
This is expected and has been observed in previous studies \citep{chevalier2006state,colburn2011state, illingworth2018estimating, amaral2021resolvent,arun2023towards,ying2023resolvent}.
The big spike of the red line in figure \ref{fig_error_corr_nonlin_lin}(a) corresponds to the large magnitudes of the estimated velocity from LE (figure \ref{fig_xzyz_nonlinear_lin}(c,d)). 
As for NE, it has a lower error and higher correlation compared with LE, indicating that NE outperforms LE. 
As there is only one measurement plane provided to NE, the nonlinear forcing is only accurate near the measurement plane. 
Nevertheless, NE with partial knowledge of the nonlinear forcing outperforms LE. 

For a statistically stationary flow, the Reynolds-Orr equation tells us that production $P$ and dissipation $D$ achieve a balance when considered over the entire domain $\Omega$:
\begin{equation}
\int_{\Omega} \frac{\mathrm{D}}{\mathrm{D}t} \left( \frac{1}{2}u_i u_i \right) \ddd \Omega =
\underbrace{\int_{\Omega} -u_i u_j \frac{\partial U_i}{\partial x_j} \ddd \Omega }_{\text{production}} + 
\underbrace{\int_{\Omega}- \frac{1}{Re}\frac{\partial u_i}{\partial x_j}\frac{\partial u_i}{\partial x_j} \ddd \Omega }_{\text{dissipation}}
\label{eqa_chapter6_Reynolds_Orr}
\end{equation}
The energy source for turbulence is production, and the energy sink for turbulence is dissipation. 
We can inspect the production and dissipation for NE and LE. 

\begin{figure}[H]
\centering
\includegraphics[scale=0.33]{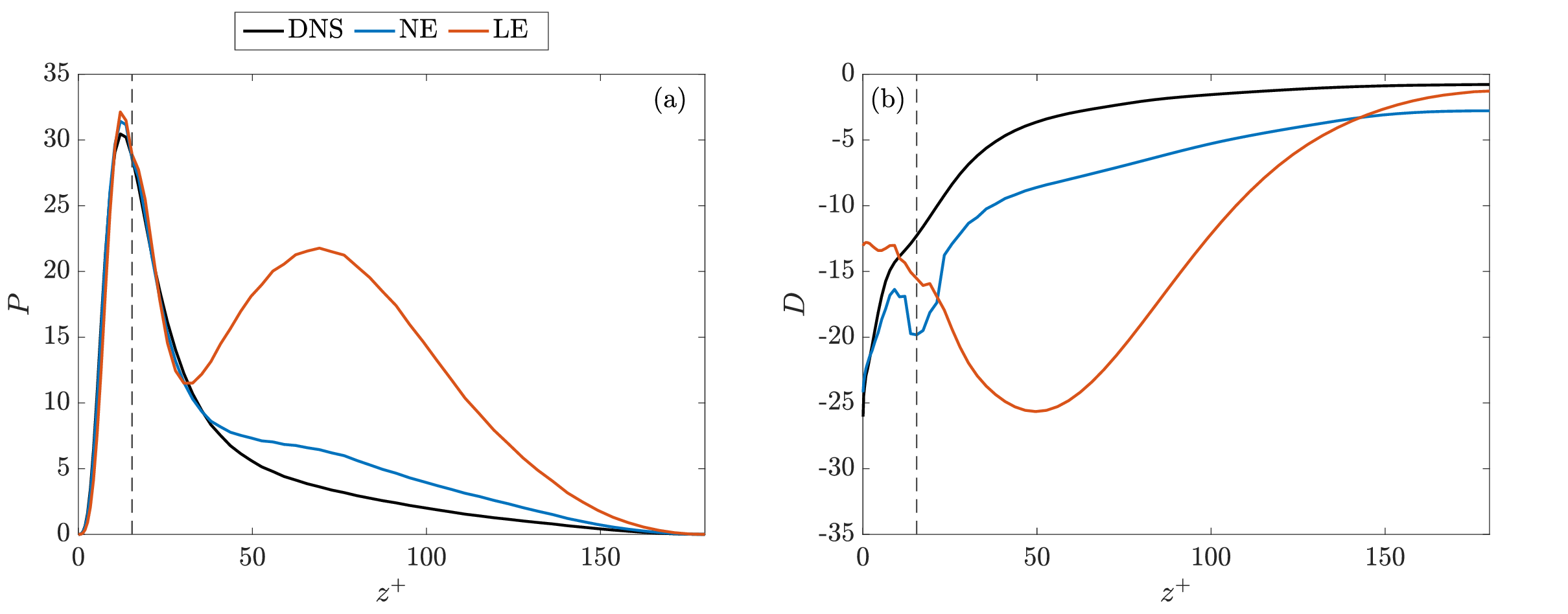}
\caption[Production and dissipation across the wall-normal height from NE and LE. ]
{(a) Production $P$ and (b) dissipation $D$ across the wall-normal height. 
The black dashed line marks the measurement plane location $z^+=15$.
}
\label{fig_P_D_NE_LE_180}
\end{figure}

Figure \ref{fig_P_D_NE_LE_180} shows the production and dissipation as a function of wall-normal height.
At each wall-normal height, the production and dissipation are averaged in the $xy$ plane and in time. 
We see good agreement for production and dissipation between NE and DNS, except for the significant dissipation near the measurement plane (the blue plunge in figure \ref{fig_P_D_NE_LE_180}(b)).
It is suspected that the sudden change of dissipation from NE is caused by the sudden change of the velocities at the measurement plane since only one measurement plane data is provided.
The sudden change of the velocities at the measurement plane will cause large velocity gradients $\frac{\partial u_i}{\partial x_j}$ contributing to the dissipation. 
For LE, substantial production and dissipation occur around $z^+=60$, corresponding to the large magnitudes of the estimated velocities (figures \ref{fig_xzyz_nonlinear_lin}(c,d)).
We can further check the sum of production and dissipation across the wall-normal height, as shown in table \ref{table_PD_NE_LE}.
According to the normalisation of \eqref{eqa_chapter6_NSEs}, the energy transfer is normalised by $u_{\tau}^3 / h$.

\setlength{\tabcolsep}{8pt} % Default value: 6pt
\renewcommand{\arraystretch}{1.2} % Default value: 1

\begin{table}[h]
\caption[Wall-normal integrated production and dissipation for NE and LE. ]
{Sum of energy transfer across the channel height of NE and LE.}
\label{table_PD_NE_LE}
\centering
\begin{tabular}{llll}
\hline
    &  $\int\;P\;\ddd z$     &    $\int\;D\;\ddd z$    &    $\int\;(P+D)\;\ddd z$   \\
DNS & 5.31  & -3.82  & 1.49  \\
NE  & 6.36  & -7.47  & -1.11 \\
LE  & 12.07  & -12.99  & -0.92 \\ \hline
\end{tabular}
\end{table}

For the DNS, the sum of production and dissipation is greater than zero because only the large scales are considered in the estimation and the energy transfer is calculated for those large scales only. 
Since small scales are mainly responsible for dissipation, we expect that the dissipation calculated from the large scales will not counterbalance production.  
Compared with the DNS, we see that both NE and LE give excessive production and dissipation.
The fact that the sum of production and dissipation for NE and LE is smaller than zero indicates that energy dissipates excessively in NE and LE. 
We do not expect that NE and LE conserve energy since the estimated velocities from the Kalman filters do not satisfy the Navier-Stokes equations because of the observation noise and the difference between estimation and observations \citep{wang2021state}.

\subsection{Comparison between NE and LEe at $Re_{\tau}=180$} \label{chapter6_section_NE_LEe_180}
In the previous section, we see that NE outperforms LE, suggesting that closing the loop of the linear resolvent-based estimator in NE is helpful. 
As an alternative way of considering the nonlinear forcing, we can use LE augmented with eddy viscosity which considers the nonlinear effects of the small scales on the large scales \citep{del2006linear,pujals2009note,illingworth2018estimating, madhusudanan2019coherent, towne2020resolvent,amaral2021resolvent, ying2023resolvent}. 
We start by performing a triple decomposition of the instantaneous velocity field, $\mathcal{U}_i = U_i + u_i + u_i^{\prime}$, where $\mathcal{U}$, $U$, $u$ and $u^{\prime}$ represent the instantaneous velocity, time-averaged velocity, large-scale organised motion and small-scale fluctuation velocity, respectively. 
The nonlinear effect of the small scales on the large-scale organised motion is modelled by an eddy viscosity \citep{reynolds1972mechanics}.
The equations of the large-scale organised motion are:
\begin{subequations}
\begin{align}
\frac{\partial u_i}{\partial t} &=  
- u_j\frac{\partial U_i}{\partial x_j} 
- U_j\frac{\partial u_i}{\partial x_j} 
-\frac{\partial p}{\partial x_i} 
+ \frac{1}{Re_{\tau}} \frac{\nu_T}{\nu} \frac{\partial^2 u_i}{\partial x_j \partial x_j}
+ f_i \\ 
\frac{\partial u_i}{\partial x_i} &= 0  
\end{align}
\label{eqa_chapter6_NSEs_eddy}%
\end{subequations}
with the eddy viscosity profile given by \citep{cess1958survey}: 
\begin{equation}
\nu_T(z) = \frac{\nu}{2}(1+\frac{\kappa^2 Re_{\tau}^2}{9}(1-z^2)^2(1+2z^2)^2[1-\mathrm{exp}(\frac{(\lvert z \rvert)Re_{\tau}}{A})]^2)^{\frac{1}{2}} + \frac{\nu}{2},
\;\;
z \in [-1,1]
\label{eqa_eddyviscosity}
\end{equation}
where $\kappa = 0.426$ and $A=25.4$, as they were the optimal values as a result of a least-square fit to experimental data at $Re_{\tau}=2000$ \citep{del2006linear}. 
The detailed expressions for the Orr-Sommerfeld and Squire operators are described in Appendix \ref{chapter6_section_appendix_OSS_state_space_model}.

\begin{figure}[H]
\centering
\includegraphics[scale=0.35]{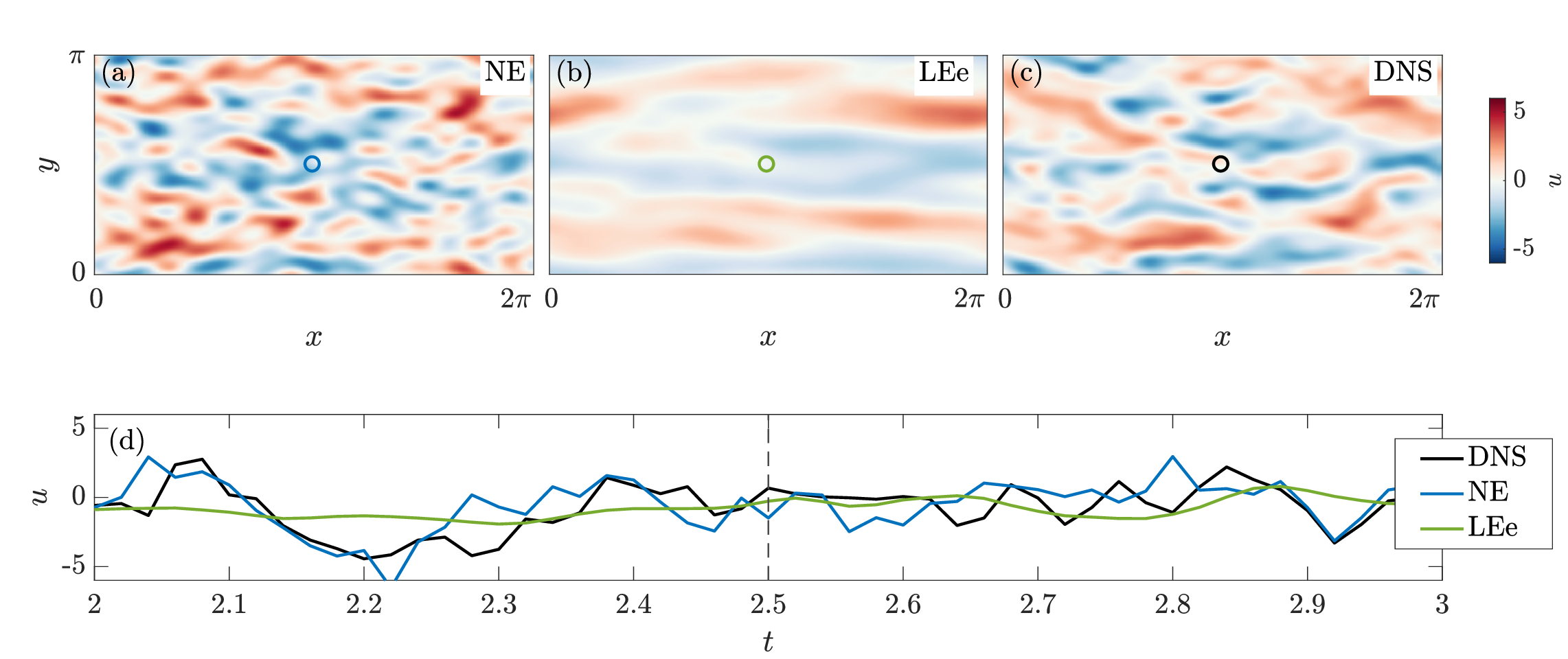}
\caption[Estimated streamwise fluctuation velocities at $z^+=50$ for NE and LEe at $Re_{\tau}=180$.]
{Streamwise velocity $u$ from (a) NE, (b) LEe and (c) DNS data in the $xy$ plane at $z^+=50$. 
The measurement data come from $z^+=15$.
The circles in the middle of (a,b,c) denote the locations of the streamwise velocity plotted in (d). 
(d) Streamwise velocity varying with time at the centre of the $xy$ plane at $z^+=50$.
The vertical dashed line denotes the time instant for (a,b,c). 
}
\label{fig_z50xy_nonlin_lin_eddy}
\end{figure}

\begin{figure}[H]
\centering
\includegraphics[scale=0.36]{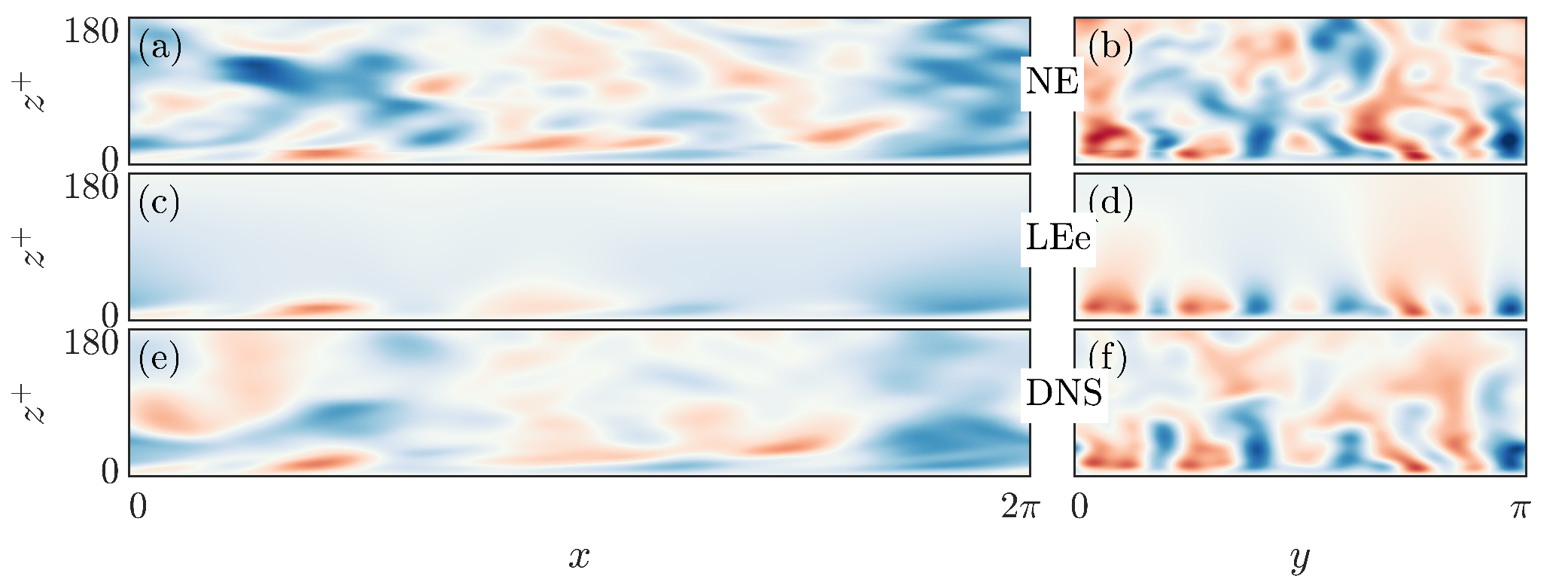}
\caption[2-D estimated streamwise velocities for NE and LEe in the $xz$ plane and $yz$ plane at $Re_{\tau}=180$.]
{(a,c,e) Streamwise velocity field in the $xz$ plane at $y=\frac{\pi}{2}$.
(b,d,f) Streamwise velocity field in the $yz$ plane at $x=\pi$. 
(a,b) NE;
(c,d) LEe and 
(e,f) the DNS data.
The limits of the colour map are the same as those in figure \ref{fig_xzyz_nonlinear_lin}.
}
\label{fig_xzyz_nonlinear_lin_eddy}
\end{figure}

Figure \ref{fig_z50xy_nonlin_lin_eddy} compares the estimated velocity fields between NE and LEe (linear estimator augmented with eddy viscosity) at $z^+=50$. 
We see that the velocity magnitudes from LEe are similar to those from the DNS and
LEe can give the correct large streamwise scales but small scales are missing (figure \ref{fig_xzyz_nonlinear_lin_eddy}(b,c)). 
Figure \ref{fig_z50xy_nonlin_lin_eddy}(d) further indicates that NE and LEe can give velocity magnitudes comparable to the DNS. 

\begin{figure}[H]
\centering
\includegraphics[scale=0.36]{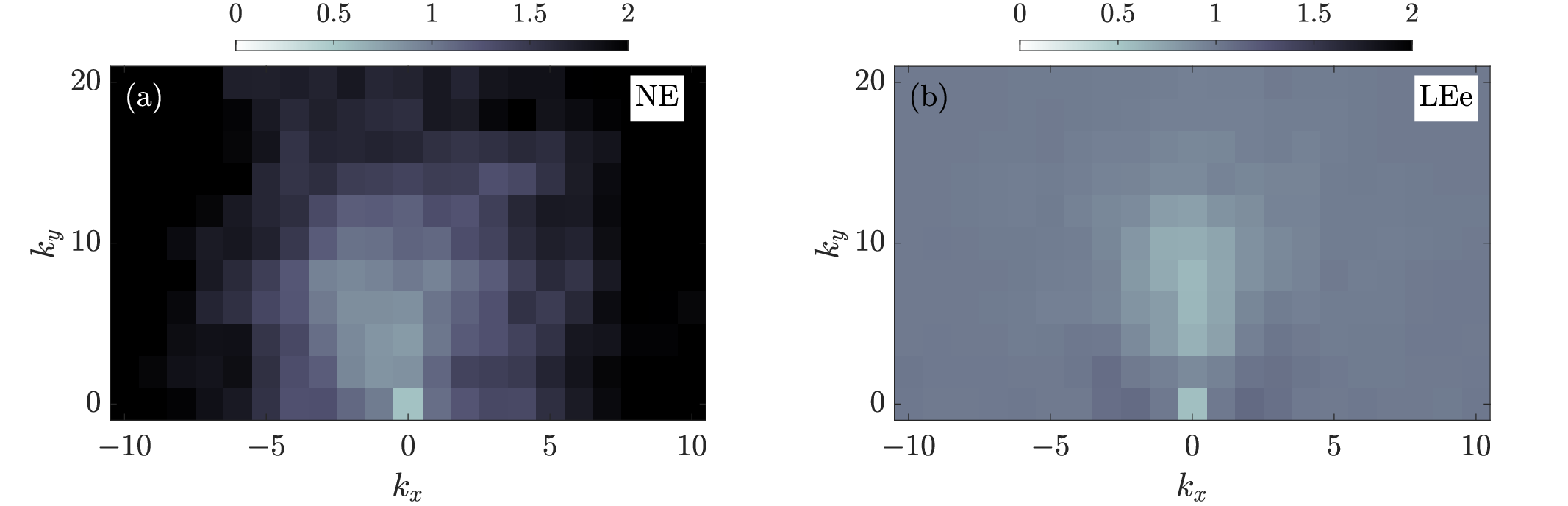}
\caption[Estimation error in Fourier space at $z^+=50$ for NE and LEe at $Re_{\tau}=180$. ]
{Error in Fourier space \eqref{eqa_error_F} for 
(a) NE and 
(b) LEe.
}
\label{fig_z50error_nonlin_lin_eddy}
\end{figure}

Figure \ref{fig_z50error_nonlin_lin_eddy} shows the error in Fourier space for NE and LEe at $z^+=50$. 
Both models perform well for the large streamwise scales with small $k_x$.
From LE to LEe, we see an improvement in estimation performance at large streamwise scales, as noted in previous work \citep{illingworth2018estimating}.
This indicates that the eddy viscosity, which considers the nonlinear effect of the small scales on the large scales, improves estimation performance. 
However, both NE and LEe have relatively large errors at small streamwise scales with large $k_x$. 
The reasons are different: for NE, it gives the incorrect small scales because only partial knowledge of the nonlinear forcing around the measurement plane is accurate (figure \ref{fig_z50xy_nonlin_lin_eddy}(a)); for LEe, it gives zeros for the small scales because LEe identifies a lack of coherence of small scales (figure \ref{fig_z50xy_nonlin_lin_eddy}(b)) \citep{madhusudanan2019coherent}.

Figure \ref{fig_xzyz_nonlinear_lin_eddy} shows the velocity field across the wall-normal height for NE and LEe. 
Both models give the velocity magnitudes and structures comparable to the DNS near the measurement plane at $z^+=15$. 
Away from $z^+=15$, the estimated velocities from NE deviate from the DNS and the estimated velocities from LEe are near zero which is noted in \citet{oehler2018linear}. 

\begin{figure}[H]
\centering
\includegraphics[scale=0.37]{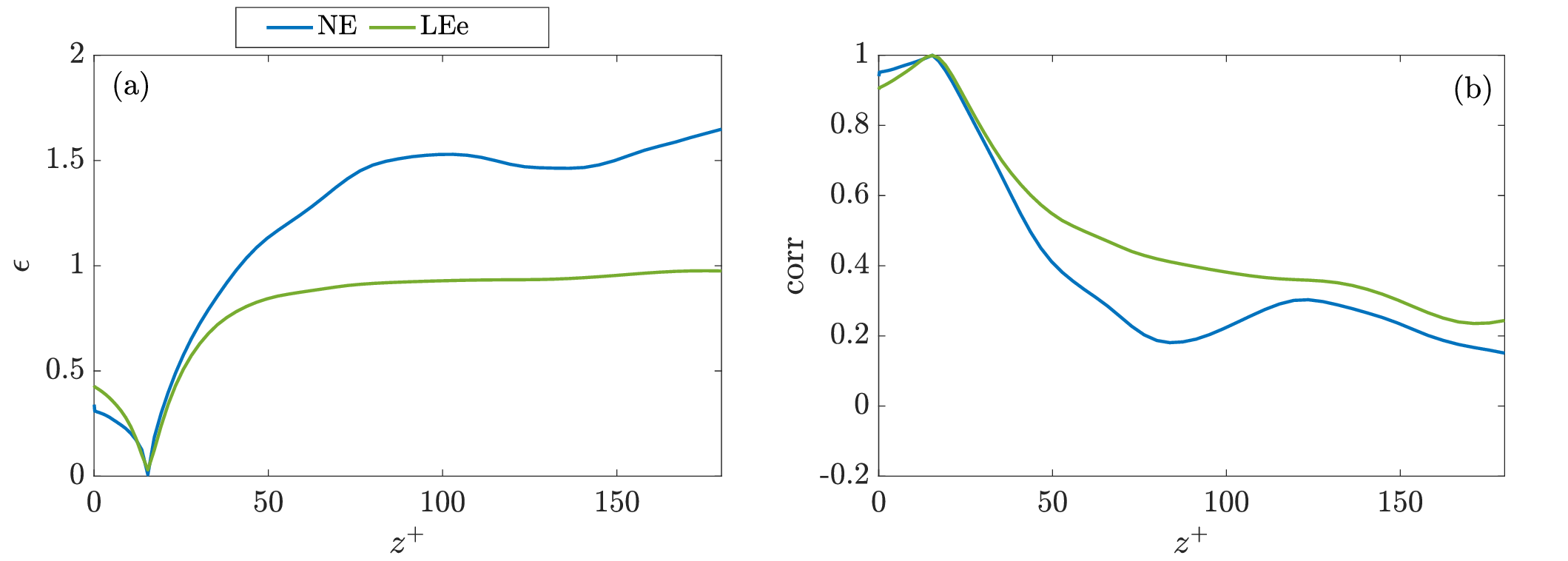}
\caption[Estimation error and correlation across the wall-normal height for NE and LEe.]
{Error \eqref{eqa_error_z} and correlation \eqref{eqa_corr_z} across the wall-normal height for NE and LEe.
}
\label{fig_error_corr_nonlin_lin_eddy}
\end{figure}

Figure \ref{fig_error_corr_nonlin_lin_eddy} quantifies the estimation performance for NE and LEe across the wall-normal height. 
Both models perform best at $z^+=15$ where the measurement data is provided, and the performance deteriorates away from the measurement plane. 
Looking at the error and correlation across the wall-normal direction, we see that LEe outperforms NE except in the near-wall region. 

We see that both NE and LEe outperform LE, indicating that considering the nonlinear forcing is important.
The difference between NE and LE is the treatment of the nonlinear forcing in the state-space model \eqref{eqa_state_space}.  
The nonlinear forcing in NE is structured since it is calculated from the estimated velocities, while the nonlinear forcing in LE is unstructured \citep{mckeon2010critical, illingworth2018estimating}. 
It has been shown that the structured nonlinear forcing gives more accurate flow statistics even though only partial knowledge of the nonlinear forcing is available \citep{zare2017colour}.
The difference between LEe and LE is the Orr-Sommerfeld and Squire operators in the state-space model \eqref{eqa_state_space}. 
Although LEe treats the nonlinear forcing as unstructured \citep{illingworth2018estimating}, the nonlinear effect of the small scales on the large scales is embedded in the state-space model \eqref{eqa_state_space} \citep{del2006linear, pujals2009note}.

\subsection{The effect of changing the measurement plane location on nonlinear estimation at $Re_{\tau}=180$} \label{chapter6_section_change_measurement_location}
%In the previous sections, the measurement plane is located at $z^+=15$ and the estimation performance deteriorates significantly farther away from the wall.
Previous studies show that using wall-shear stress as measurement only gives good velocity estimation up to the logarithmic region \citep{chevalier2006state,colburn2011state,suzuki2017estimation, amaral2021resolvent} and is no better than using velocity as measurement \citep{oehler2018linear}. 
Since we investigate using the velocity measurement only at $z^+=15$ in the previous sections, we would like to know the effect of changing the measurement plane location in the wall-normal direction, especially farther away from the wall. 
%This section investigates the effect of changing the measurement plane location in the wall-normal direction. 
We change the measurement plane location from $z^+=15$ to $50,75,100,125$ and $150$, respectively. 
For each estimation case, we further evaluate the overall estimation performance by integrating the error and correlation in the wall-normal direction:
\begin{subequations}
\begin{gather}
\epsilon_{int} =  \int \limits_{0}^{1} \; \epsilon\; \ddd z ,   \quad   \mathrm{corr}_{int} = \int \limits_{0}^{1} \; \mathrm{corr}\; \ddd z     \tag{\theequation a,b}
\end{gather}
\label{eqa_error_corr_int}%
\end{subequations}

\begin{figure}[H]
\centering
\includegraphics[scale=0.37]{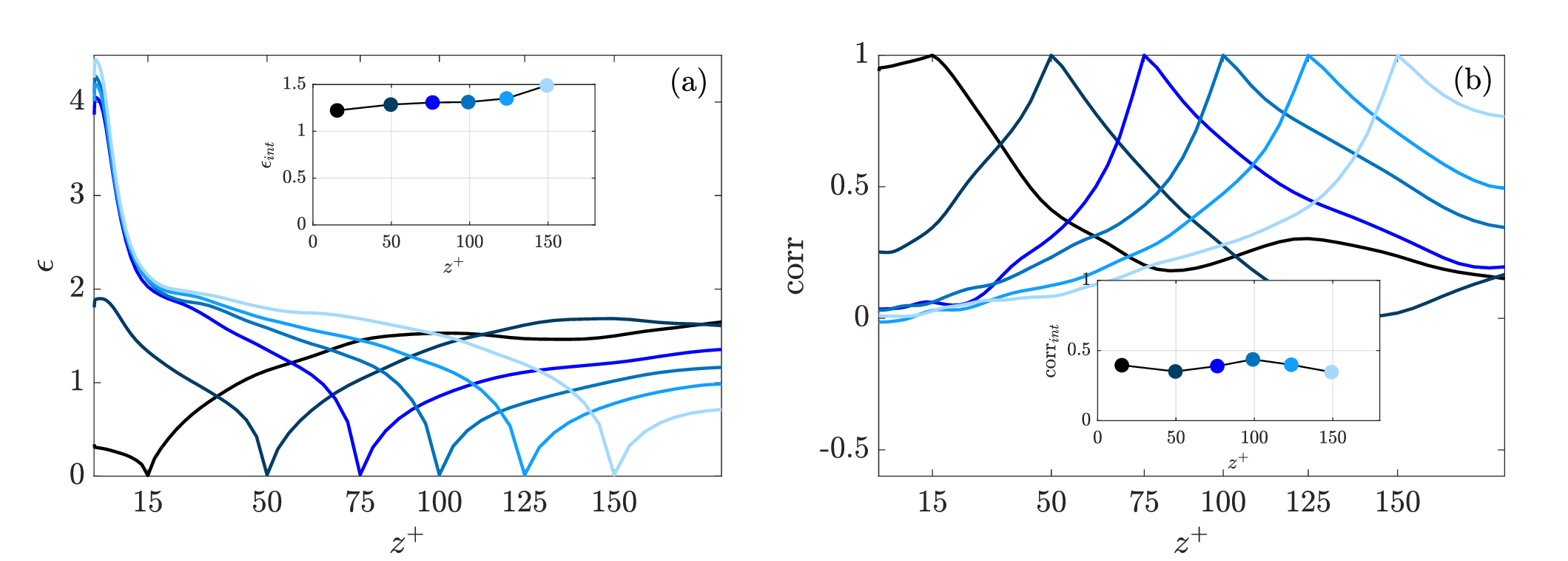}
\caption[Estimation error and correlation for NE when changing measurement location at $Re_{\tau}=180$.]
{
Error \eqref{eqa_error_z} and correlation \eqref{eqa_corr_z} for NE when the measurement plane is at $z^+=15,50,75,100,125$ and $150$, respectively. 
An estimation case with a measurement plane located farther away from the wall is indicated by a lighter blue colour. 
Inset plots show the integrated error and correlation \eqref{eqa_error_corr_int}. 
}
\label{fig_error_corr_different_measurement_locations}
\end{figure}

Figure \ref{fig_error_corr_different_measurement_locations} shows the estimation results with the measurement plane at different wall-normal heights. 
No matter where we put the measurement plane, the estimation performance deteriorates away from the measurement location and the overall performance does not change substantially as seen from the inset plots. 
When the measurement plane is placed farther away from the wall $(z^+=75,100,125,150)$, the error in the near-wall region increases significantly (figure \ref{fig_error_corr_different_measurement_locations}(a)).
The estimated statistics in the near-wall region could be improved by embedding the relationship between the inner layer and outer layer \citep{marusic2010predictive, baars2016spectral}. 

\subsection{Nonlinear estimation at $Re_{\tau}=590$} \label{chapter6_section_estimation_590_one_plane}
\begin{figure}[H]
\centering
\includegraphics[scale=0.45]{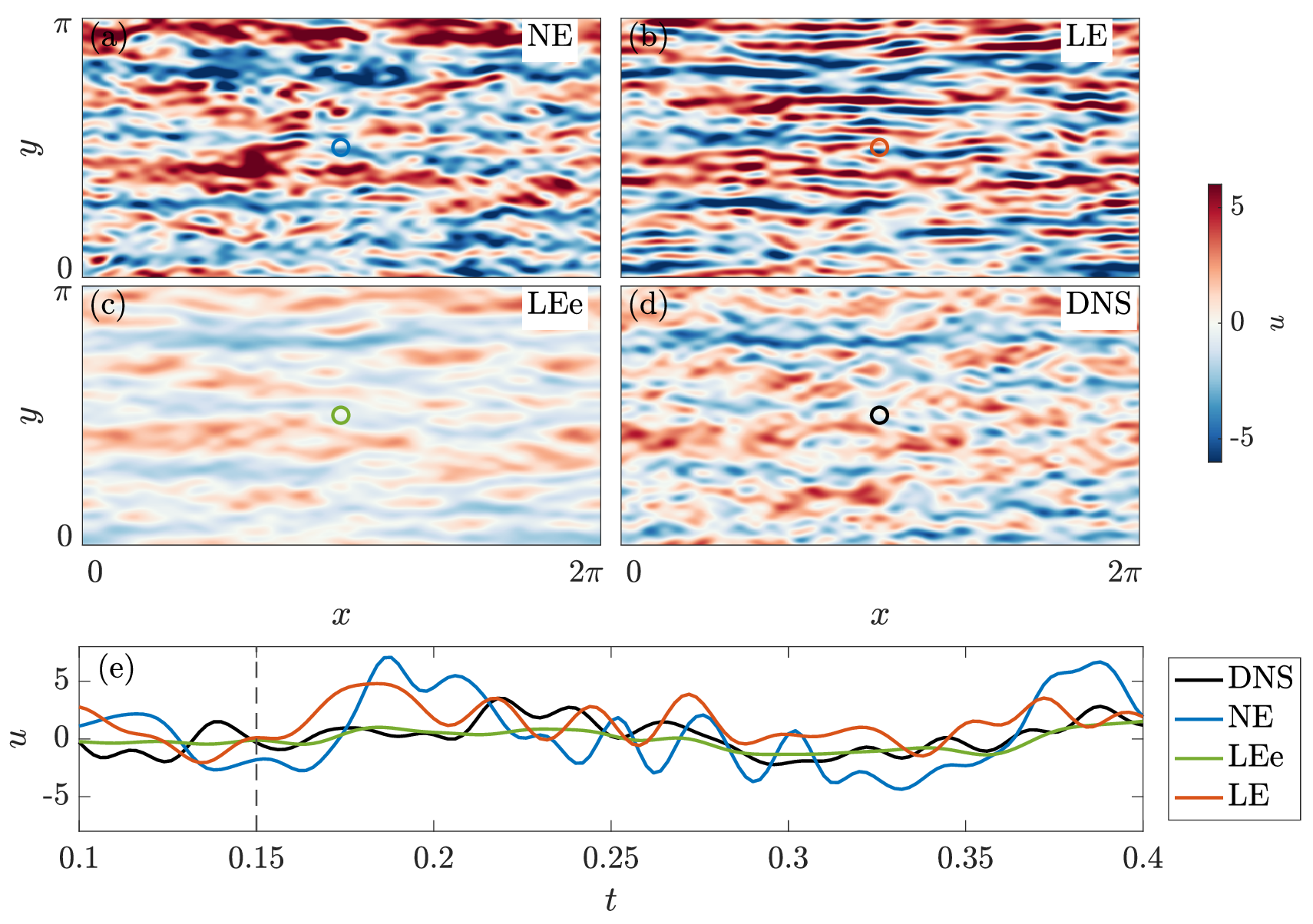}
\caption[Estimated streamwise fluctuation velocities at $z^+=50$ for NE, LE and LEe at $Re_{\tau}=590$.]
{    
Streamwise velocity $u$ from (a) NE, (b) LE, (c) LEe and (d) DNS data in the $xy$ plane at $z^+=50$ and $Re_{\tau}=590$.
The measurement data come from $z^+=18$.
The circles in the middle of (a,b,c,d) denote the locations of the streamwise velocity plotted in (e).
(e) Streamwise velocity varying with time at the centre of the $xy$ plane at $z^+=50$.
The vertical dashed line denotes the time instant for (a,b,c,d).
}
\label{fig_z50_xy_590}
\end{figure}

Channel flow estimation has been investigated at high Reynolds numbers \citep{illingworth2018estimating, oehler2018linear,amaral2021resolvent,wang2022observable,ying2023resolvent} and we wonder if the nonlinear estimator designed in this study can be applied at a higher Reynolds number. 
In this section, we apply the nonlinear estimator at $Re_{\tau}=590$. 
The wavenumbers considered at $Re_{\tau}=590$ are $\lvert k_x \rvert \leq 20$ and $\lvert k_y \rvert \leq 40$, corresponding to structures with $\lambda_x^+ \geq 185$ and $\lambda_y^+ \geq 93$.
In the wall-normal direction we use 257 Chebyshev points and convergence has been checked by doubling the number of Chebyshev points.
The measurement data (velocities) come from $z^+=18$, where the turbulent kinetic energy reaches its maximum in the wall-normal direction. 
The measurement data comes from DNS and the temporal resolution is $\Delta T_m = 0.002(\Delta T_m^+=1.18)$ which is small enough to resolve the frequency of the smallest scale according to Taylor's hypothesis. 
The SSROM simulation time step is set to be $\Delta T_s = 0.00005$.
The estimation starts with zero initial condition. 
We present the estimation results starting from $t=0.1(t^+=59)$ to $t=0.4(t^+=236)$ to minimise the initial transient.

Figure \ref{fig_z50_xy_590} shows the estimation results using different estimators at $Re_{\tau}=590$. 
From the colour scale in figures \ref{fig_z50_xy_590}(a,b,c,d), we see that both NE and LE overpredict the velocity magnitudes; LEe underpredicts the velocity magnitudes. 
From the shapes of the structures, we see that NE tends to break the large-scale structures presented in the DNS into small-scale structures; LE only gives structures that are long in the streamwise direction and thin in the spanwise direction; LEe retains the shapes of the large-scale structures. 
Figure \ref{fig_z50_xy_590}(e) further shows that the velocity from NE deviates from the DNS substantially. 

\begin{figure}[H]
\centering
\includegraphics[scale=0.35]{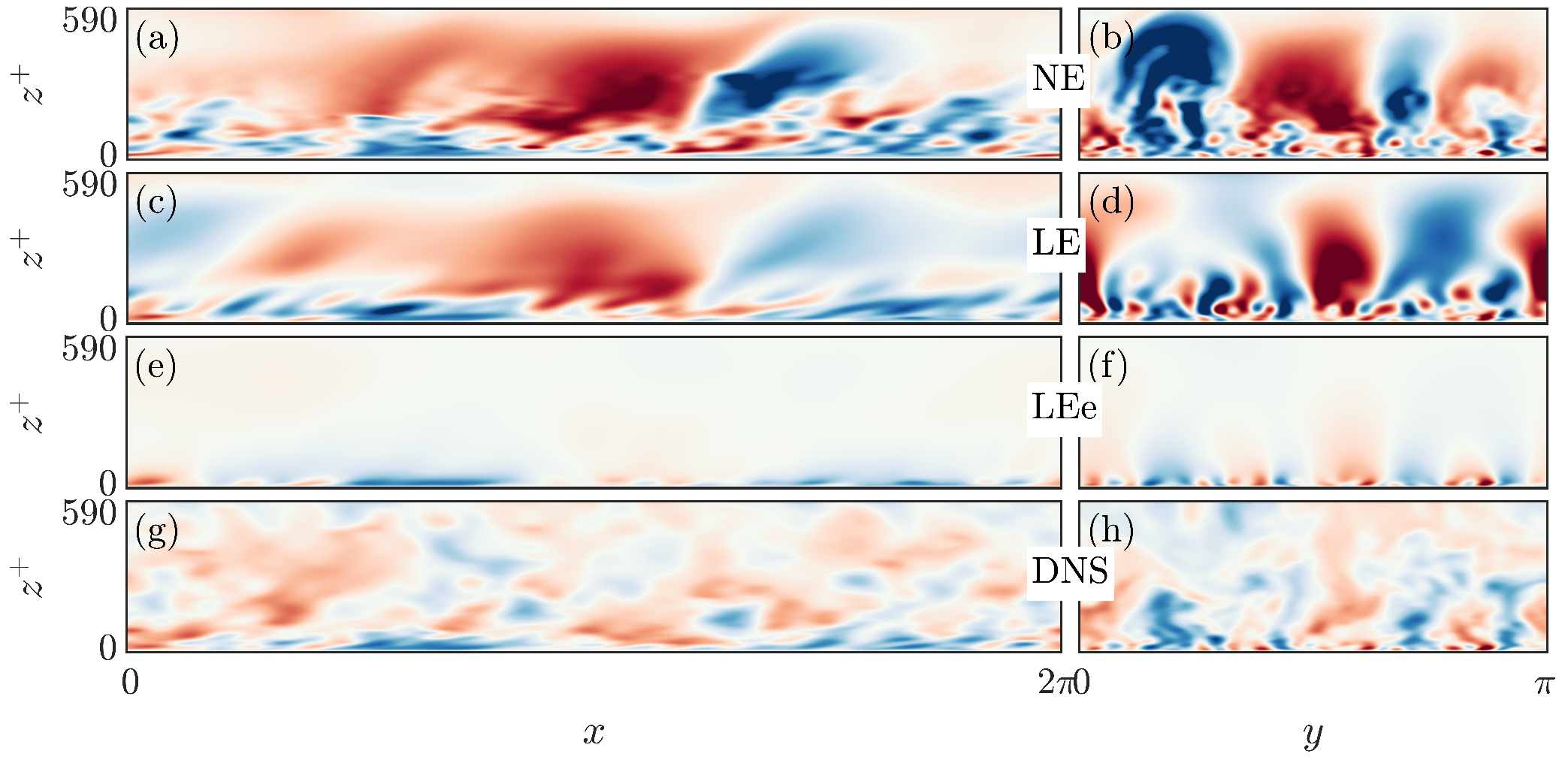}
\caption[2-D estimated streamwise velocities for NE, LE and LEe in the $xz$ plane and $yz$ plane at $Re_{\tau}=590$.]
{    
(a,c,e,g) Streamwise velocity field in the $xz$ plane at $y = \frac{\pi}{2}$.
(b,d,f,h) Streamwise velocity field in the $yz$ plane at $x=\pi$.
(a,b) NE;
(c,d) LE;
(e,f) LEe and 
(g,h) the DNS data. 
The limits of the colour map are the same as that in figure \ref{fig_xzyz_590}.
}
\label{fig_xzyz_590}
\end{figure}

From the previous section, we see that estimators using one measurement plane do not provide a good overall result across the channel height at $Re_{\tau}=180$, irrespective of the location of the measurement plane. 
Therefore, we do not expect that estimators using only one measurement plane will work well across the channel height at $Re_{\tau}=590$. 
Figure \ref{fig_xzyz_590} shows the streamwise velocity fields across the wall-normal height. 
We see that all estimators cannot give the correct fields away from the near-wall region (the measurement plane is at $z^+=18$). 
NE and LE overpredict the values; the velocity magnitudes outside the near-wall region from LEe are approximately zero, which is also noted in \citet{oehler2018linear}. 
Further including small scales in the estimation does not show noticeable improvement. 
For the estimation with only one measurement plane provided, NE does not perform well.
LEe performs well near the measurement plane and gives zero velocities at wall-normal heights that do not share coherence with the measurement plane, indicating the effectiveness of the eddy viscosity at higher Reynolds numbers \citep{illingworth2018estimating,oehler2018linear}.

\section{Nonlinear estimation using multiple measurement planes}
\label{chapter6_section_results_multiple_planes}
We have seen from the previous section that the performance of the nonlinear estimator deteriorates significantly away from the measurement plane with only one measurement plane provided. 
Due to the inhomogeneity caused by the wall in wall-bounded flows, the flow physics varies significantly in the wall-normal direction.
It has been shown that the coherence between two wall-normal heights decreases as the two wall-normal heights are separated farther \citep{madhusudanan2019coherent}. 
This implies that even though we consider the effect of the nonlinear forcing in NE and LEe, the underlying physics forbids us from estimating the velocities that are not coherent with the measurement data \citep{illingworth2018estimating}. 
The failure to estimate the velocities away from the measurement plane can also be attributed to insufficient measurement data in the wall-normal direction.
The OSS state-space model \eqref{eqa_state_space} contains a sufficient number of states (129 Chebyshev points at $Re_{\tau}=180$) distributed in the buffer layer, log layer and outer layer in the wall-normal direction in order to describe the flow statistics comprehensively.
Therefore, the nonlinear estimator with measurement data at only one wall-normal height is not able to estimate the velocities across the channel.

\citet{arun2023towards} investigated channel flow estimation using multiple measurement planes in the wall-normal direction.
Inspired by this, we increase the number of measurement planes provided to the nonlinear estimator. 
Specifically, \S\ref{chapter6_section_multiple_planes_180} discusses the results at $Re_{\tau}=180$ and \S\ref{chapter6_section_multiple_planes_590} discusses the results at $Re_{\tau}=590$.
%We further relate the nonlinear estimation using multiple measurement planes to large-eddy simulation (LES).

\subsection{At $Re_{\tau}=180$}
\label{chapter6_section_multiple_planes_180}
We test four cases with the number of measurement planes $N_{mea}$ varying from 1 to 4. 
The details are listed in table \ref{table_multiple_measurment}.

\setlength{\tabcolsep}{8pt} % Default value: 6pt
\renewcommand{\arraystretch}{1.2} % Default value: 1

\begin{table}[h]
\caption{
Wall-normal locations of the measurement planes for four cases.
}
\label{table_multiple_measurment}
\centering
\begin{tabular}{lllll}
\hline
$N_{mea}$ & 1  & 2     & 3         & 4             \\ 
$z^+_{mea}$ & 15 & 15,50 & 15,50,100 & 15,50,100,150 \\ \hline
\end{tabular}
\end{table}

\begin{figure}[H]
\centering
\includegraphics[scale=0.37]{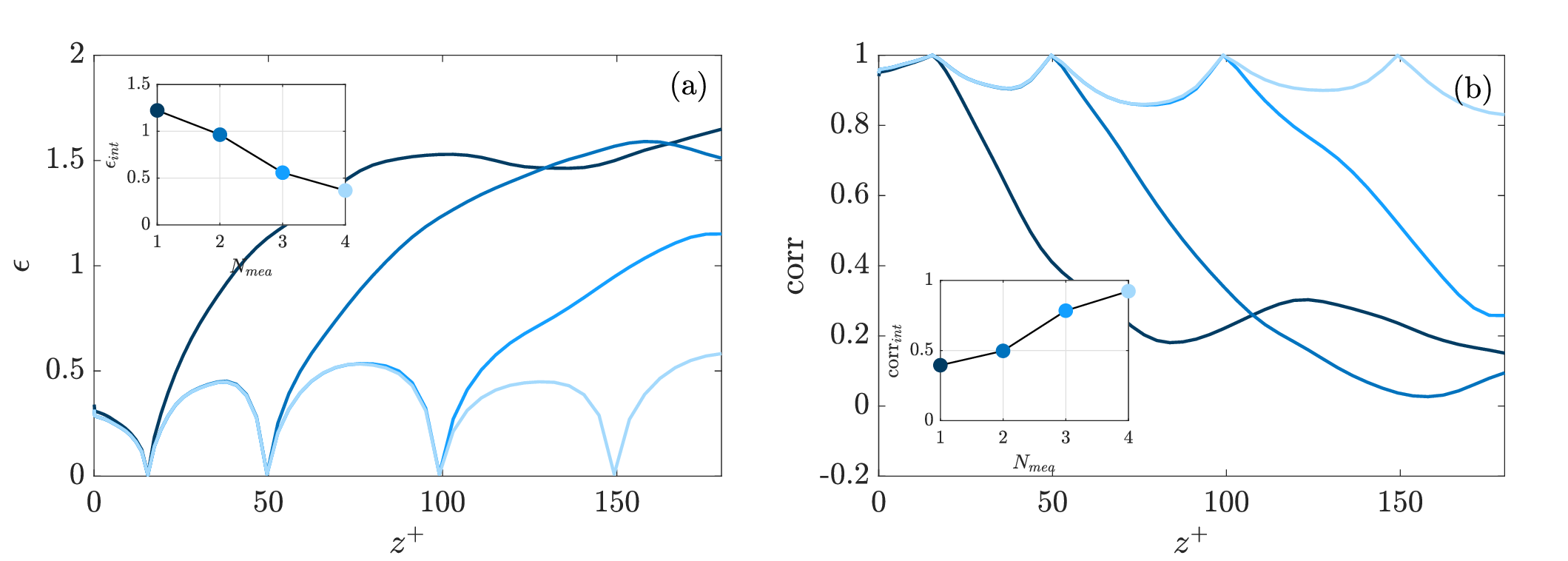}
\caption[Estimation error and correlation for NE when measurements at multiple wall-normal heights are available at $Re_{\tau}=180$.]
{    
Error \eqref{eqa_error_z} and correlation \eqref{eqa_corr_z} for NE with multiple measurement planes considered as listed in table \ref{table_multiple_measurment}.
A case with a larger number of measurement planes is shown using a lighter blue colour.  
Inset plots show the integrated error and correlation \eqref{eqa_error_corr_int}.}
\label{fig_error_corr_multiple_measurement_planes}
\end{figure}

Figure \ref{fig_error_corr_multiple_measurement_planes} shows the estimation performance of the four cases listed in table \ref{table_multiple_measurment}. 
As expected, NE gives the correct velocities at the measurement locations \citep{arun2023towards}. 
However, even if the number of measurement planes increases, substantial error still occurs when the estimation plane is far from the measurement plane, as seen in the cases with $N_{mea}=1,2,3$.

As for the overall estimation performance indicated by $\epsilon_{int}$ and $\mathrm{corr}_{int}$ \eqref{eqa_error_corr_int}, we see that the overall estimation significantly improves with the inclusion of just a few more measurement planes (inset plots in figure \ref{fig_error_corr_multiple_measurement_planes}). 
If the measurement planes are more evenly distributed in the wall-normal direction (for example the case with $N_{mea}=4$), the error and correlation do not deteriorate significantly (the curves flatten out).
In terms of achieving a good overall estimation performance, an open question concerns the number of measurement planes needed and the distribution of the measurement planes in the wall-normal direction. 
The visualisation of the estimated velocity fields for NE with multiple measurements are shown for the $Re_{\tau}=590$ case in \S\ref{chapter6_section_multiple_planes_590} and are not shown for the $Re_{\tau}=180$ case because the results are similar.

\subsection{At $Re_{\tau}=590$} \label{chapter6_section_multiple_planes_590}
\begin{figure}[H]
\centering
\includegraphics[scale=0.45]{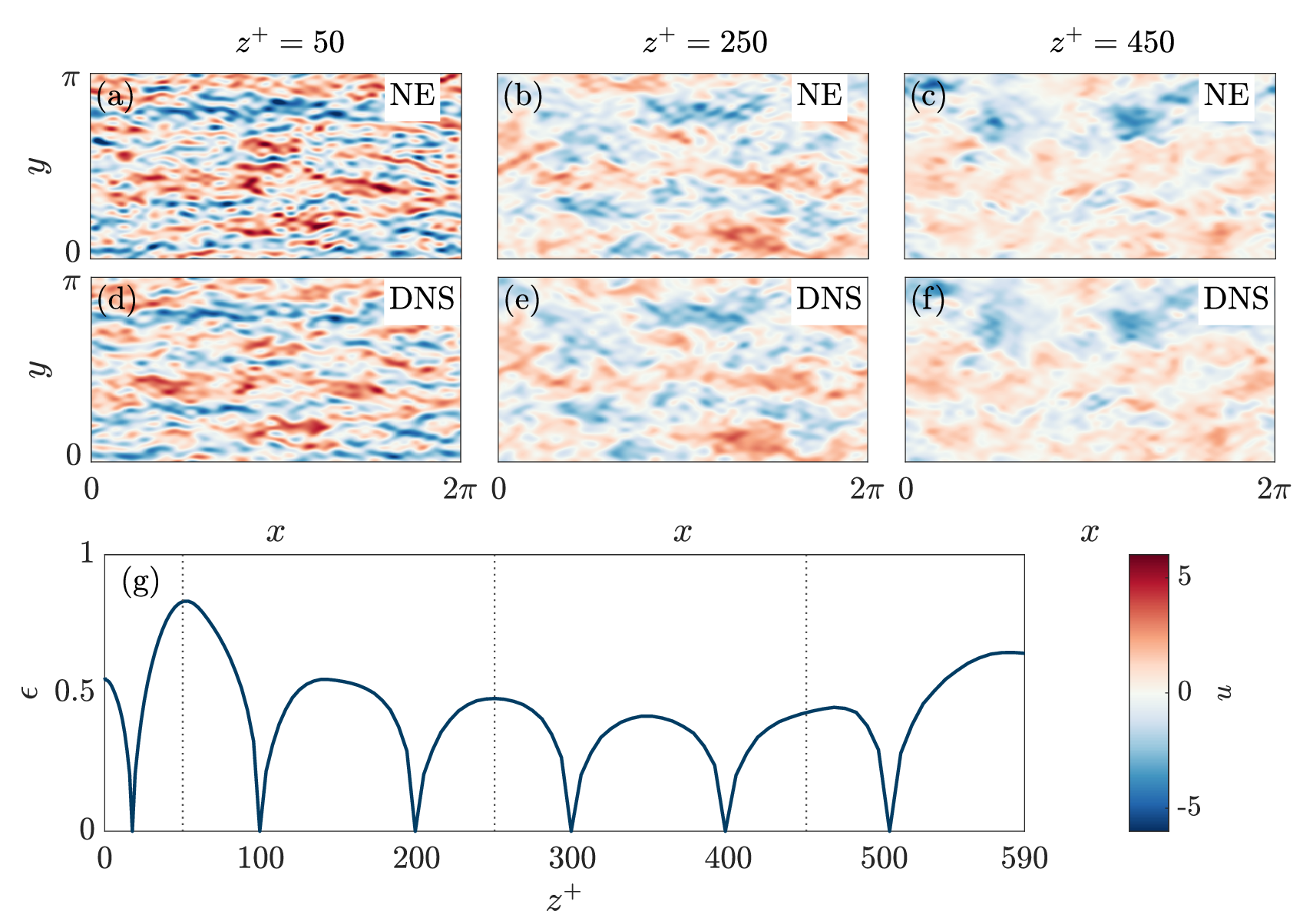}
\caption[Nonlinear estimation with six multiple measurement planes.]
{
Results in $xy$ plane of nonlinear estimation with six measurement planes at $z^+ = 18, 100,200,300,400$ and $500$.
(a,b,c) The estimated velocity fields for NE;
(d,e,f) The velocity fields from DNS. 
(a,d) $z^+=50$; (b,e) $z^+=250$; (c,f) $z^+=450$.
(g) Error \eqref{eqa_error_z} for NE. Three dotted lines mark $z^+ = 50, 250$ and $450$.
}
\label{fig_zmultiple_xy_590}
\end{figure}

From \S\ref{chapter6_section_multiple_planes_180}, we know that multiple measurement planes evenly distributed across the wall-normal height enhances the NE performance significantly. 
Therefore, we increase the number of measurement planes from one to six at $Re_{\tau}=590$, with measurement planes at $z^+ = 18, 100,200,300,400$ and $500$. 
Figure \ref{fig_zmultiple_xy_590} shows the estimated velocity fields at $z^+ = 50, 250, 450$ and the estimation error across the channel height (the correlation is similar to the error and is not shown). 
As expected, figure \ref{fig_zmultiple_xy_590}(g) shows that NE gives accurate estimation results at the measurement planes.
At the three chosen wall-normal heights $z^+=50,250$ and $450$ (figure \ref{fig_zmultiple_xy_590}(a-f)), NE gives satisfactory estimation results for large-scale structures, though NE slightly overpredicts the velocity magnitudes at a few locations. 

\begin{figure}[H]
\centering
\includegraphics[scale=0.37]{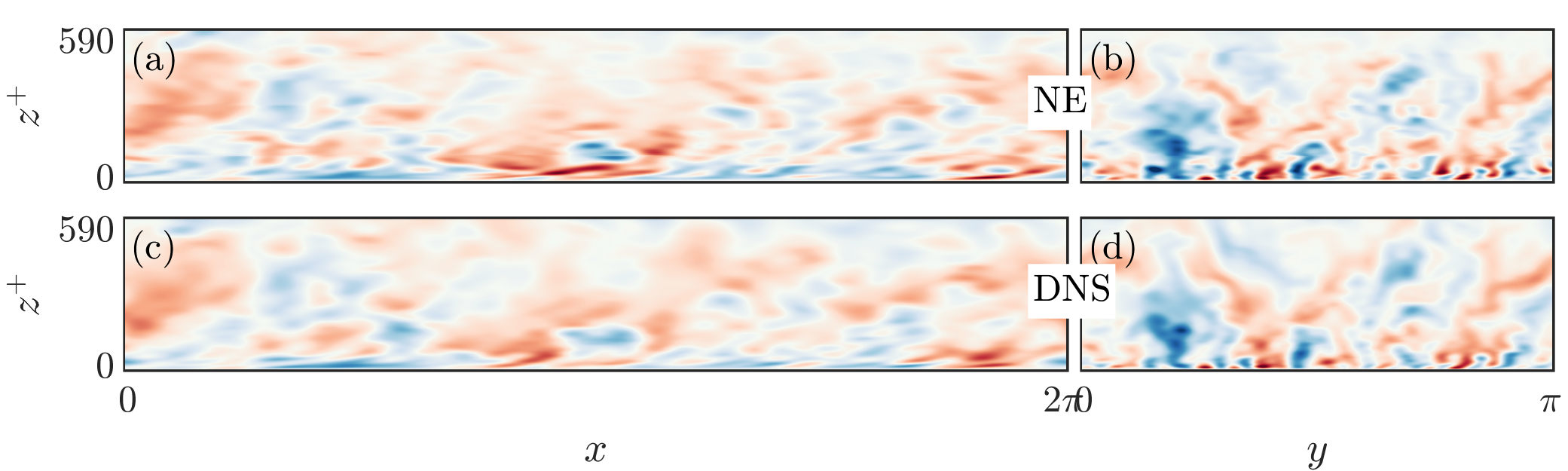}
\caption[Nonlinear estimation with six multiple measurement planes.]
{
Results in $xz$ and $yz$ planes of nonlinear estimation with six measurement planes at $z^+ = 18, 100,200,300,400$ and $500$.
(a,b) The estimated velocity fields for NE;
(c,d) The velocity fields from DNS. 
(a,c) $xz$ plane; (b,d) $yz$ plane.
}
\label{fig_multiple_xzyz_590}
\end{figure}

Figure \ref{fig_multiple_xzyz_590} shows the estimated velocity fields across the channel height. 
Compared with the single measurement plane case in figure \ref{fig_xzyz_590}, increasing the number of measurement planes greatly improves the estimation performance across the channel height. 
Even with the increased number of measurement planes, this number is still substantially smaller than the number of DNS gridpoints in the wall-normal direction.
% As mentioned in \S\ref{chapter6_section_multiple_planes_180}, NE with multiple measurement planes across the channel height can be linked to LES, where coarser grids are distributed in the wall-normal direction. 

The nonlinear estimator in this study can be related to large-eddy simulation (LES).
Both LES and nonlinear estimation only consider the largest scales. 
The main difference is that LES is autonomous, whereas the nonlinear estimator requires external information continuously supplied (the measurement data). 
In LES, the energy transfer between the resolved scales and subgrid scales is modelled to improve accuracy.
In light of this and the energy transfer in figure \ref{fig_P_D_NE_LE_180}, a possible way to improve the nonlinear estimator would be to consider the energy transfer.

\section{Conclusions} \label{chapter6_section_conclusions}
We investigate turbulent channel flow estimation at $Re_{\tau}=180$ and $590$.
The nonlinear estimator designed in this study consists of two main parts. 
The linear part is based on the linearised Navier-Stokes equations with nonlinear forcing as input and velocity as output \citep{illingworth2018estimating}.
%The linear part is implemented in Fourier space. 
The nonlinear part links the velocity back to the nonlinear forcing through the nonlinear dynamics $f_i = - u_j \frac{\partial u_i}{\partial x_j}$.
%The nonlinear part is implemented in physical space. 
%We only consider the energetically dominant Fourier modes corresponding to the largest scales.
%The estimation is implemented at a measurement sampling time $\Delta T_m$.
%To avoid divergence, we embed the state-space reduced-order (SSROM) simulation running at a much smaller time step $\Delta T_s$ between two consecutive Kalman filter updates determined by $\Delta T_m$.

We compare the estimation performance between the nonlinear estimator (NE), linear estimator (LE) and linear estimator augmented with eddy viscosity (LEe).
At $Re_{\tau}=180$, NE and LEe outperform LE in terms of estimating the velocity magnitudes, structure shapes and energy transfer across the wall-normal height.
%By inspecting the estimation performance in Fourier space, we see that NE and LEe have lower estimation errors for large streamwise scales than LE in which the large streamwise scales are excessively amplified.
%In terms of energy transfer, we see that NE and LE give excessive production and dissipation, while LEe does not give sufficient production and dissipation.
We see that the overall estimation performance does not change substantially as the location of the measurement plane is varied.
%and will improve significantly if multiple measurement planes are evenly distributed in the wall-normal direction. 
At $Re_{\tau}=590$, NE does not work well using only one measurement plane. 
The performance of the nonlinear estimator can be significantly enhanced by including a few more measurement planes evenly distributed in the wall-normal direction.
The nonlinear estimator in this study shares similarities with large-eddy simulation (LES).

The linear estimator augmented with eddy viscosity outperforms the nonlinear estimator designed in this study in terms of the estimation performance, stability, simplicity and applicability to high Reynolds number flows. 
This has shown the usefulness of eddy viscosity although eddy viscosity only serves to dissipate energy. 
Apart from using the nonlinear estimation methodology in this study and eddy viscosity, there are other ways to consider the nonlinearity, such as using extended Kalman filter \citep{chevalier2006state}, ensemble Kalman filter \citep{colburn2011state}, inner-outer layer relationship \citep{baars2016spectral}, cross-spectral density (CSD) tensor \citep{towne2020resolvent} and so on.
A promising future work would be to combine the advantages of different nonlinear estimation strategies. 

\backmatter

\bmhead{Acknowledgements}
This research was supported by The University of Melbourne’s Research Computing Services and the Petascale Campus Initiative.

\section*{Declarations}

\begin{itemize}
\item Conflict of interest: The authors declare that they have no conﬂict of interest.
\item Data availability: The datasets generated and analysed during the current study are available from the corresponding
author on reasonable request.
\end{itemize}

\newpage

\begin{appendices}

\section{State-space form of the linear model} \label{chapter6_section_appendix_OSS_state_space_model}
The detailed expressions for matrices $\mathcal{A}$, $\mathcal{B}$ and $\mathcal{C}$ of the state-space model \eqref{eqa_state_space} are:
\begin{subequations}
\begin{align}
\mathcal{A} &= 
\begin{bmatrix}
\Delta^{-1}\mathcal{L}_{OS} & 0 \\
-\mathrm{i}k_y\frac{\ddd U}{\ddd z} & \mathcal{L}_{SQ}
\end{bmatrix} \\ 
\mathcal{B} &= 
\begin{bmatrix}
-\mathrm{i}k_x\Delta^{-1}\mathnormal{D} & -\mathrm{i}k_y\Delta^{-1}\mathnormal{D} & -k^2\Delta^{-1} \\
-\mathrm{i}k_y & -\mathrm{i}k_x & 0
\end{bmatrix} \\
\mathcal{C} &= 
\frac{1}{k^2}
\begin{bmatrix}
\mathrm{i}k_x\mathnormal{D} & -\mathrm{i}k_y\\
\mathrm{i}k_y\mathnormal{D} & \mathrm{i}k_x\\
k^2 & 0
\end{bmatrix}
\end{align}
\label{eqa_ABC}%
\end{subequations}
with boundary conditions $\boldsymbol{\hat{w}}(t)=\frac{\partial \hat{\bm{w}}}{\partial z}(t)=\boldsymbol{\hat{\eta}}(t)=0$ at the two walls.
$\mathrm{i} = \sqrt{-1}$ is the imaginary unit.
$\mathnormal{D}$ is the differentiation matrix in the wall-normal direction; 
$k^2= k_x^2+k_y^2$; 
$\Delta = \mathcal{D}^2-k^2$ is the Laplacian operator; 
and $\mathcal{L}_{OS}$ and $\mathcal{L}_{SQ}$ are the Orr-Sommerfeld and Squire operators, respectively:
\begin{subequations}
\begin{align}
\mathcal{L}_{OS} &= -\mathrm{i} k_x U \Delta + \mathrm{i} k_x \frac{\ddd^2 U}{\ddd z^2} + \frac{1}{Re_{\tau}}\Delta^2  \label{eqa_OS}
\\
\mathcal{L}_{SQ} &= 
-\mathrm{i} k_x U  + \frac{1}{Re_{\tau}}\Delta \label{eqa_SQ}
\end{align}
\end{subequations}

The operators are discretised using the Chebyshev polynomial in the wall-normal direction \citep{trefethen2000spectral}. 
The boundary conditions at the walls are implemented following \citet{trefethen2000spectral, weideman2000matlab}.
The integration in the wall-normal direction is implemented using Clenshaw-Curtis quadrature \citep{trefethen2000spectral}.

The linear model \eqref{eqa_state_space} is a building block for the linear and nonlinear estimators.

\begin{figure}[H]
\centering
\includegraphics[scale=0.50]{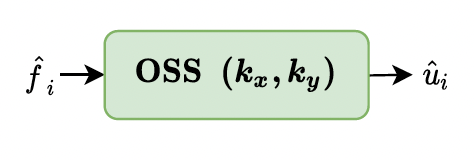}
\caption[The block diagram of the Orr-Sommerfeld \& Squire state-space model.]
{The Orr-Sommerfeld \& Squire state-space model \eqref{eqa_state_space} maps the nonlinear forcing $\hat{f}_i$ to the velocities $\hat{u}_i$ at a single wavenumber pair $(k_x,k_y)$ in Fourier space.}
\label{fig_OSS_state_space}
\end{figure}

For the linear model with eddy viscosity \eqref{eqa_chapter6_NSEs_eddy}, $\mathcal{L}_{OS}$ and $\mathcal{L}_{SQ}$ change to:
\begin{subequations}
\begin{align}
\mathcal{L}_{OS} &= -\mathrm{i} k_x U \Delta + \mathrm{i} k_x \frac{\ddd^2 U}{\ddd z^2} + \frac{1}{Re_{\tau}}\frac{\nu_T}{\nu} \Delta^2 + 2\frac{1}{Re_{\tau}}\frac{1}{\nu}\frac{\ddd \nu_T}{\ddd z} \mathnormal{D} \Delta + \frac{1}{Re_{\tau}}\frac{1}{\nu}\frac{\ddd^2 \nu_T}{\ddd z^2}(\mathnormal{D}^2 + k^2)
\label{eqa_OS_eddy}
\\
\mathcal{L}_{SQ} &= 
-\mathrm{i} k_x U  + \frac{1}{Re_{\tau}} \frac{\nu_T}{\nu} \Delta  + \frac{1}{Re_{\tau}} \frac{1}{\nu} \frac{\ddd \nu_T}{\ddd z} \mathnormal{D} \label{eqa_SQ_eddy}%
\end{align}
\end{subequations}

\section{More details about the linear estimator} \label{chapter6_section_appendix_linear_estimator}
\begin{figure}[H]
\centering
\includegraphics[scale=0.50]{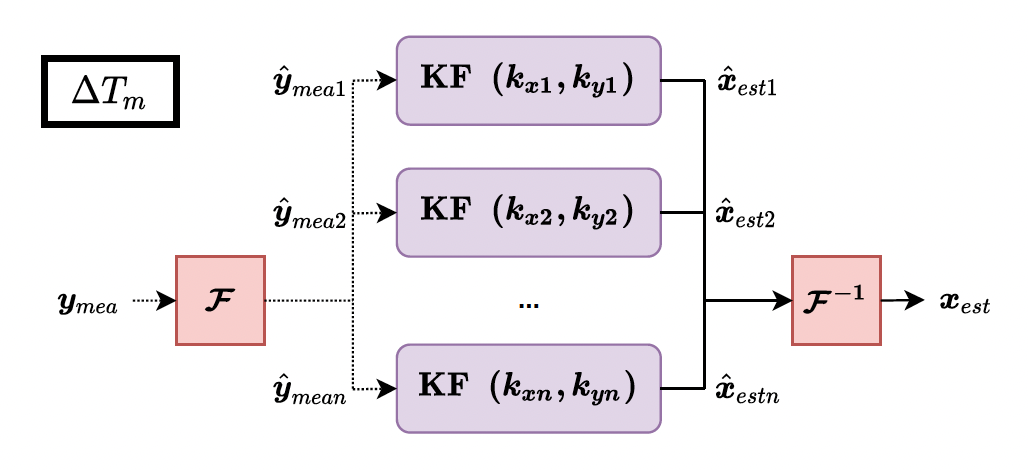}
\caption[The block diagram of the linear estimator.]
{The structure of the linear estimator corresponding to the diagram in figure \ref{fig_simple_linear_estimator}.
$\mathcal{F}$ represents the Fourier transform and
 $\mathcal{F}^{-1}$ represents the inverse Fourier transform.
Each Kalman filter (KF) receives the measurement $\hat{\bm{y}}_{mea}$ and outputs the estimated full state $\hat{\bm{x}}_{est}$.
Each Kalman filter is implemented in Fourier space at measurement sampling time $\Delta T_m$.
}
\label{fig_linear_estimator}
\end{figure}

The linear estimator \eqref{eqa_estimated_state} is implemented at one wavenumber pair $(k_x,k_y)$. 
For the wavenumber range considered, multiple Kalman filters at different wavenumber pairs are implemented in parallel.
Algorithm \ref{algorithm_linear_estimation} details the processes of the linear estimation and the structure of the linear estimator is illustrated in figure \ref{fig_linear_estimator}. 
Since estimation processes at different wavenumbers are independent, parallel programming could be used to accelerate the computation \citep{kepner2009parallel}.
\begin{algorithm}[H]
\small
\setstretch{0.95}
\SetKwInput{KwInput}{Input}                % Set the Input
\SetKwInput{KwOutput}{Output}              % set the Output
\DontPrintSemicolon
  
  \KwInput{measurement data $y_{mea}$, considered wavenumber pairs, measurement sampling time $\Delta T_m$}
  \KwOutput{estimated velocities $\hat{x}_{est}$}
  %\KwData{Testing set $x$}

% Set Function Names
  \SetKwFunction{FMain}{Main}
  \SetKwFunction{FSum}{Sum}
  \SetKwFunction{FSub}{Sub}
 
% Write Function with word ``Function''
  \SetKwProg{Fn}{Kalman filter estimation (every measurement sampling time $\Delta T_m$)}{:}{}
  \Fn{}{
        obtain the measurement data $y_{mea}$ in physical space \;
        take the 2-D Fourier transform to get $\hat{y}_{mea}$ in wavenumber space \;
        \SetKwProg{Fn}{for}{do}{}
        \Fn{every $(k_x,k_y)$ considered $\;$}{
              use the Kalman filter to get the estimated velocities $\hat{x}_{est}$ in wavenumber space: 
              \begin{equation*}
                  \frac{\ddd}{\ddd t} \hat{\bm{x}}_{est} = \mathcal{A} \hat{\bm{x}}_{est}(t) + 
\mathcal{L} [\hat{\bm{y}}_{mea}(t) - \mathcal{C}_{mea} \hat{\bm{x}}_{est}(t)] 
              \end{equation*}
              \tcc{the marching time step for Kalman filter is $\Delta T_m$ }
             }
        take the inverse 2-D Fourier transform to get $x_{est}$ in physical space\;

  }
\caption{Linear Estimation}
\label{algorithm_linear_estimation}
\end{algorithm}

\section{More details about the nonlinear estimator} \label{chapter6_section_appendix_nonlinear_estimator}
\begin{figure}[H]
\centering
\includegraphics[scale=0.42]{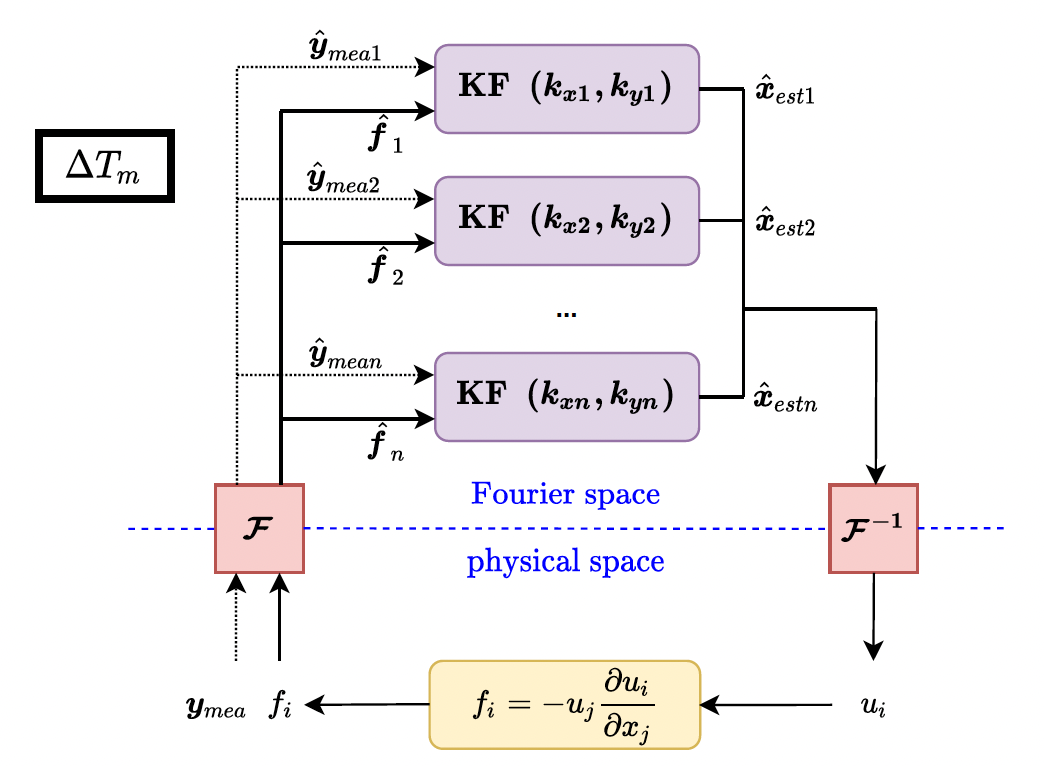}
\caption[The block diagram of the nonlinear estimator.]
{The structure of the nonlinear estimator corresponding to the diagram in figure \ref{fig_simple_nonlinear_estimator1}.
For each Kalman filter at a single wavenumber pair $(k_x,k_y)$, it receives the nonlinear forcing $\hat{\bm{f}}$ and measurement data $\hat{\bm{y}}_{mea}$, and outputs the estimated full velocity state $\hat{\bm{x}}_{est}$.
The nonlinear forcing is explicitly calculated from the previously estimated velocities in physical space. 
Each Kalman filter is implemented in Fourier space at measurement sampling time $\Delta T_m$.
}
\label{fig_nonlinear_estimator1}
\end{figure}

%The main idea of the nonlinear estimator is that we close the loop of the linear estimator in figure \ref{fig_linear_estimator} by considering the nonlinear dynamics in physical space $f_i = -u_j \frac{\partial u_i}{\partial x_j}$. 
Algorithm \ref{algorithm_nonlinear_estimation} details the processes of the nonlinear estimation and the structure of the nonlinear estimator is illustrated in figure \ref{fig_nonlinear_estimator1}.

\begin{algorithm}[H]
\small
\setstretch{0.95}
\SetKwInput{KwInput}{Input}                % Set the Input
\SetKwInput{KwOutput}{Output}              % set the Output
\DontPrintSemicolon
  
  \KwInput{measurement data $y_{mea}$, considered wavenumber pairs, measurement sampling time $\Delta T_m$}
  \KwOutput{estimated velocities $\hat{x}_{est}$}
  %\KwData{Testing set $x$}

% Set Function Names
  \SetKwFunction{FMain}{Main}
  \SetKwFunction{FSum}{Sum}
  \SetKwFunction{FSub}{Sub}
 
% Write Function with word ``Function''
  \SetKwProg{Fn}{Kalman filter estimation (every measurement sampling time $\Delta T_m$)}{:}{}
  \Fn{}{
        obtain the measurement data $y_{mea}$ and nonlinear forcing $f_i$ in physical space \;
        take the 2-D Fourier transform to get $\hat{y}_{mea}$ and $\hat{f}_i$ in wavenumber space \;
        \SetKwProg{Fn}{for}{do}{}
        \Fn{every $(k_x,k_y)$ considered $\;$}{
              use the Kalman filter to get the estimated velocities $\hat{x}_{est}$ in wavenumber space: 
              \begin{equation*}
                  \frac{\ddd}{\ddd t} \hat{\bm{x}}_{est} = \mathcal{A} \hat{\bm{x}}_{est}(t) +
\mathcal{B} \hat{\bm{f}}(t) + 
\mathcal{L} [\hat{\bm{y}}_{mea}(t) - \mathcal{C}_{mea} \hat{\bm{x}}_{est}(t)] 
              \end{equation*}
              \tcc{the marching time step for Kalman filter is $\Delta T_m$ }
             }
        take the inverse 2-D Fourier transform to get $x_{est}$ in physical space\;
        use the nonlinear dynamic to get the nonlinear forcing $f_i$ in physical space:
        \begin{equation*}
            f_i = -u_j \frac{\partial u_i}{\partial x_j}
        \end{equation*}
        
  }
\caption{Nonlinear Estimation}
\label{algorithm_nonlinear_estimation}
\end{algorithm}

\section{State-space reduced-order model simulation} \label{chapter6_section_appendix_SSROM}
\begin{figure}[H]
\centering
\includegraphics[scale=0.50]{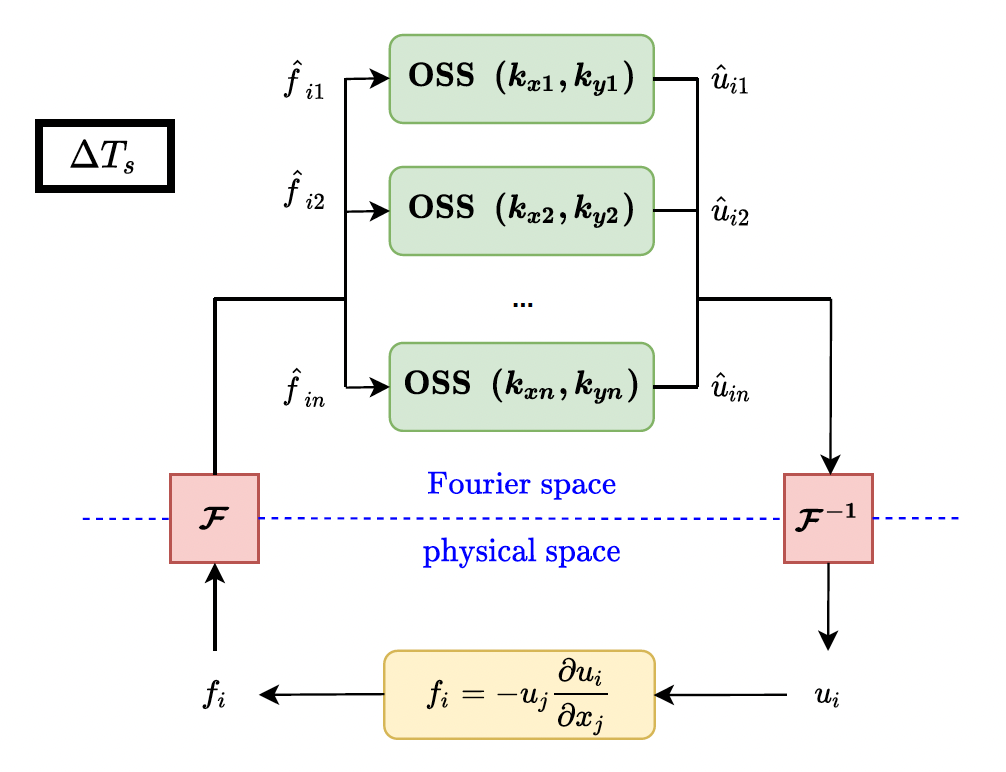}
\caption[The block diagram of state-space reduced-order model (SSROM) simulation.]
{The structure of state-space reduced-order model (SSROM) simulation.
The upper part above the blue dashed line is implemented in Fourier space and the lower part below the blue dashed line is implemented in physical space.
The SSROM simulation is implemented at the simulation time step $\Delta T_s$.
}
\label{fig_ssrom}
\end{figure}

The state-space reduced-order model (SSROM) serves for autonomous simulations and is a key component of the revised nonlinear estimator \ref{fig_simple_nonlinear_estimator2}.
The basic idea is to use nonlinear forcing $\hat{f}_i$ to drive the OSS state-space model in figure \ref{fig_OSS_state_space} to obtain velocities $\hat{u}_i$ in Fourier space and to use the nonlinear dynamics $f_i= -u_j \frac{\partial u_i}{\partial x_j}$ to link the velocities $u_i$ back to nonlinear forcing $f_i$ in physical space. 
To comply with the estimation setting in which we only consider the large scales, this autonomous simulation model only considers the same large scales as well. 
Since the total number of wavenumber pairs considered is much smaller than that considered in a DNS, we call this autonomous simulation model `state-space reduced-order model (SSROM) simulation'. 
Algorithm \ref{algorithm_SSROM} details the processes of the SSROM simulation and the structure of the SSROM simulation is illustrated in figure \ref{fig_ssrom}.

The stability of the SSROM simulation is determined by the simulation time step $\Delta T_s$. 
For a certain range of wavenumbers considered, $\Delta T_s$ needs to be set small enough to avoid divergence. 
In this study, $\Delta T_s$ needs to be set small enough for the SSROM simulation to `survive' the measurement sampling time $\Delta T_m$ for a smooth implementation of the nonlinear estimation. 

\begin{algorithm}[H]
\small
\setstretch{0.95}
\SetKwInput{KwInput}{Input}                % Set the Input
\SetKwInput{KwOutput}{Output}              % set the Output
\DontPrintSemicolon
  
  \KwInput{initial velocity fields $u_{i0}$, considered wavenumber pairs, SSROM simulation time step $\Delta T_s$, total simulation time $T_{end}$}
  \KwOutput{simulated velocities $u_i$}
  %\KwData{Testing set $x$}

% Set Function Names
  \SetKwFunction{FMain}{Main}
  \SetKwFunction{FSum}{Sum}
  \SetKwFunction{FSub}{Sub}
 
% Write Function with word ``Function''
        \color{Green}
        \SetKwProg{Fn}{SSROM simulation (running at $\Delta T_s$)}{:}{}
        \Fn{}{

        create a counter: $n_{SSROM}=0$ \;
        use the nonlinear dynamic to get the initial nonlinear forcing $f_{i0}$ in physical space:
        \begin{equation*}
            f_{i0} = -u_{j0} \frac{\partial u_{i0}}{\partial x_{j0}}
        \end{equation*}
        
        \SetKwProg{Fn}{while}{do}{}
        \Fn{$n_{SSROM}< \frac{T_{end}}{\Delta T_s} \;$}{

        take the 2-D Fourier transform to get the nonlinear forcing $\hat{f}_i$ in wavenumber space\;

        \SetKwProg{Fn}{for}{do}{}
        \Fn{every $(k_x,k_y)$ considered $\;$}{
              use the Orr-Sommerfeld \& Squire state-space model to get the velocity $\hat{u}_i$ in wavenumber space \\
              \tcc{the marching time step for the OSS state-space model is $\Delta T_s$ }
             }

        take the inverse 2-D Fourier transform to get the velocities $u_i$ in physical space\;
        use the nonlinear dynamic to get the nonlinear forcing $f_i$ in physical space:
        \begin{equation*}
            f_i = -u_j \frac{\partial u_i}{\partial x_j}
        \end{equation*}\;
        $n_{SSROM} = n_{SSROM}+1$
        }

        }

\caption{State-Space Reduced-Order Model (SSROM) Simulation}
\label{algorithm_SSROM}
\end{algorithm}

\section{More details about the revised nonlinear estimator} \label{chapter6_section_appendix_nonlinear_estimator_SSROM}
\begin{figure}[H]
\centering
\includegraphics[scale=0.70]{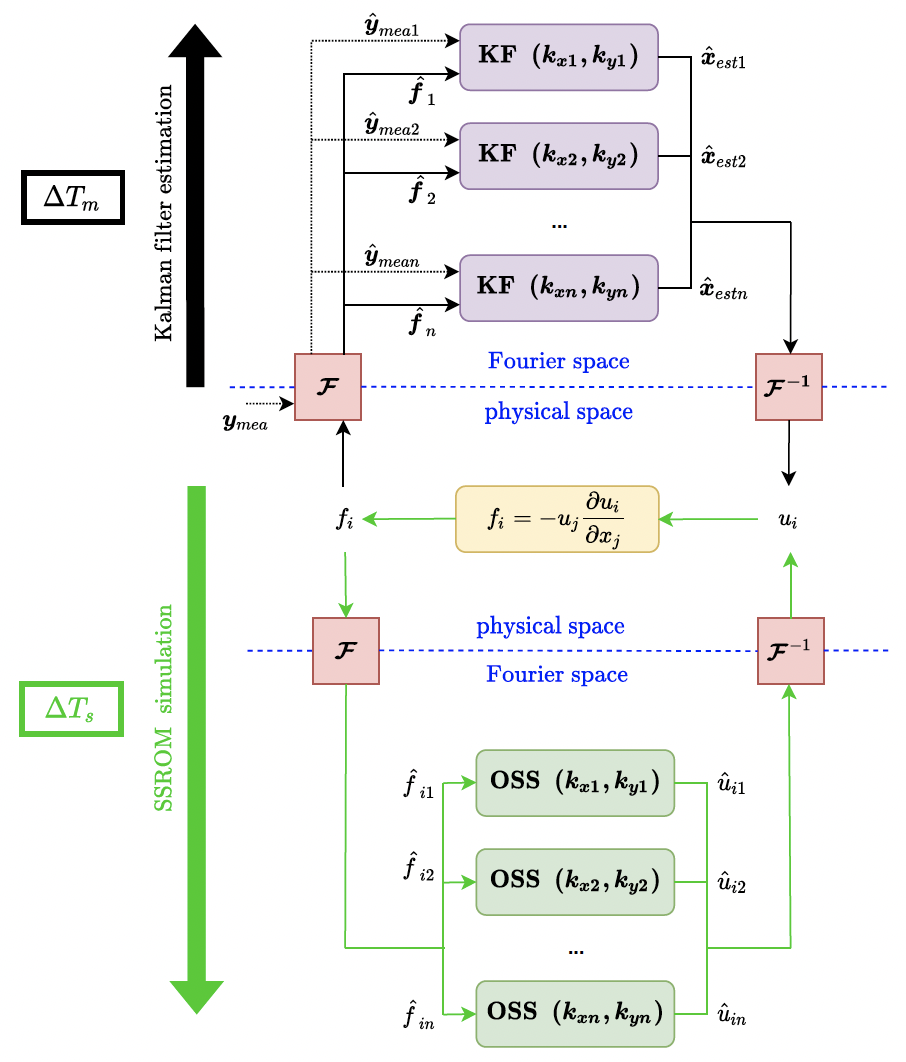}
\caption[The block diagram of the revised nonlinear estimator.]
{The structure of the revised nonlinear estimator, which is composed of the linear resolvent-based estimator \eqref{eqa_state_space_KF_for_nonlinear} and SSROM simulation. 
Black lines illustrate the Kalman filter estimation part which is the same as the model in figure \ref{fig_linear_estimator}. 
Kalman filter is used every measurement sampling time $\Delta T_m$. 
Green lines illustrate the SSROM simulation part which is implemented at the simulation time step $\Delta T_s$. 
$\Delta T_s \ll \Delta T_m$ to avoid divergence. 
}
\label{fig_nonlinear_estimator2}
\end{figure}

The nonlinear estimator illustrated in figure \ref{fig_nonlinear_estimator1} gives divergent estimation results easily if the measurement sampling time $\Delta T_m$ is not small enough. 
To overcome this problem, we embed the SSROM simulation running at a time step of $\Delta T_s \; (\Delta T_s \ll \Delta T_m)$ between two consecutive Kalman filter updates separated by measurement sampling time $\Delta T_m$.
Algorithm \ref{algorithm_nonlinear_estimation_with_SSROM} details the processes of the revised nonlinear estimation and the structure of the revised nonlinear estimator is illustrated in figure \ref{fig_nonlinear_estimator2}.

The time step for the SSROM simulation $\Delta T_s$ needs to be small enough so that the SSROM simulation can `survive' the measurement sampling time $\Delta T_m$.  
A general tip for selecting an appropriate $\Delta T_s$ is to run several SSROM simulations with different time steps and to pick out the cases without divergence. 
In addition, the divergence problem is related to how much external information is provided to the estimator. 
If the number of measurement planes increases, the occurrence of divergence will greatly decrease.

\begin{algorithm}[H]
\small
\setstretch{0.95}
\SetKwInput{KwInput}{Input}                % Set the Input
\SetKwInput{KwOutput}{Output}              % set the Output
\DontPrintSemicolon
  
  \KwInput{measurement data $y_{mea}$, considered wavenumber pairs, measurement sampling time $\Delta T_m$, SSROM simulation time step $\Delta T_s$}
  \KwOutput{estimated velocities $\hat{x}_{est}$}
  %\KwData{Testing set $x$}

% Set Function Names
  \SetKwFunction{FMain}{Main}
  \SetKwFunction{FSum}{Sum}
  \SetKwFunction{FSub}{Sub}
 
% Write Function with word ``Function''
  \SetKwProg{Fn}{Kalman filter estimation (every measurement sampling time $\Delta T_m$)}{:}{}
  \Fn{}{
        obtain the measurement data $y_{mea}$ and nonlinear forcing $f_i$ in physical space \;
        take the 2-D Fourier transform to get $\hat{y}_{mea}$ and $\hat{f}_i$ in wavenumber space \;
        \SetKwProg{Fn}{for}{do}{}
        \Fn{every $(k_x,k_y)$ considered $\;$}{
              use the Kalman filter to get the estimated velocities $\hat{x}_{est}$ in wavenumber space: 
              \begin{equation*}
                  \frac{\ddd}{\ddd t} \hat{\bm{x}}_{est} = \mathcal{A} \hat{\bm{x}}_{est}(t) +
\mathcal{B} \hat{\bm{f}}(t) + 
\mathcal{L} [\hat{\bm{y}}_{mea}(t) - \mathcal{C}_{mea} \hat{\bm{x}}_{est}(t)] 
              \end{equation*}
              \tcc{the marching time step for Kalman filter is $\Delta T_s$ and Kalman filter is used every measurement sampling time $\Delta T_m$ }
             }
        take the inverse 2-D Fourier transform to get $x_{est}$ in physical space\;
        use the nonlinear dynamic to get the nonlinear forcing $f_i$ in physical space:
        \begin{equation*}
            f_i = -u_j \frac{\partial u_i}{\partial x_j}
        \end{equation*}

        create a counter for SSROM simulation: $n_{SSROM}=0$

        \color{Green}
        \SetKwProg{Fn}{SSROM simulation (running at $\Delta T_s$)}{:}{}
        \Fn{}{

        \SetKwProg{Fn}{while}{do}{}
        \Fn{$n_{SSROM}< \frac{\Delta T_m}{\Delta T_s} \;$}{

        take the 2-D Fourier transform to get the nonlinear forcing $\hat{f}_i$ in wavenumber space\;

        \SetKwProg{Fn}{for}{do}{}
        \Fn{every $(k_x,k_y)$ considered $\;$}{
              use the Orr-Sommerfeld \& Squire state-space model to get the velocity $\hat{u}_i$ in wavenumber space\;
              \tcc{the marching time step for the OSS state-space model is $\Delta T_s$ }
             }

        take the inverse 2-D Fourier transform to get the velocities $u_i$ in physical space\;
        use the nonlinear dynamic to get the nonlinear forcing $f_i$ in physical space:
        \begin{equation*}
            f_i = -u_j \frac{\partial u_i}{\partial x_j}
        \end{equation*}\;
        $n_{SSROM} = n_{SSROM}+1$
        }

        }
        
  }
\caption{Revised Nonlinear Estimation (Kalman Filter Estimation Embedded with SSROM Simulation)}
\label{algorithm_nonlinear_estimation_with_SSROM}
\end{algorithm}

\end{appendices}

%%===========================================================================================%%
%% If you are submitting to one of the Nature Portfolio journals, using the eJP submission   %%
%% system, please include the references within the manuscript file itself. You may do this  %%
%% by copying the reference list from your .bbl file, paste it into the main manuscript .tex %%
%% file, and delete the associated \verb+\bibliography+ commands.                            %%
%%===========================================================================================%%

\bibliography{sn-bibliography}% common bib file
%% if required, the content of .bbl file can be included here once bbl is generated
%%\input sn-article.bbl

\end{document}